\documentclass[aps,pra,showpacs,amsmath,amssymb,twocolumn,superscriptaddress,notitlepage,preprintnumbers]{revtex4-2}
\usepackage{graphicx}
\usepackage{amsmath,amssymb,amsthm,mathrsfs,amsfonts,dsfont}
\usepackage{subfigure, epsfig}
\usepackage{braket}
\usepackage{physics}
\usepackage{bm}
\usepackage{enumerate}
\usepackage{color,xcolor}
\usepackage{qcircuit}
\usepackage{graphicx}
\usepackage{algorithm}
\usepackage{algpseudocode}
\usepackage{hyperref}
\usepackage[svgnames]{xcolor}
\hypersetup{
     colorlinks = true,
     linkcolor = Maroon,
     anchorcolor = black,
     citecolor = DarkBlue,
     filecolor = blue,
    urlcolor = DarkBlue
     }
\usepackage{comment}
\usepackage[normalem]{ulem}

\usepackage{varioref}
\labelformat{equation}{(#1)}

\labelformat{section}{#1}

\labelformat{figure}{#1}

\labelformat{proposition}{#1}

\labelformat{lemma}{#1}

\labelformat{theorem}{#1}

\labelformat{observation}{#1}

\labelformat{definition}{#1}

\labelformat{problem}{#1}

\labelformat{algorithm}{#1}

\labelformat{corollary}{#1}

\labelformat{statement}{#1}

\newtheorem{theorem}{Theorem}

\newtheorem{proposition}{Proposition}

\newtheorem{lemma}{Lemma}

\newtheorem{corollary}{Corollary}

\begin{document}
 
\title{Quantum computational spectroscopy of complex resonances and optical residues}
 
\date{\today}

\author{Payman Mahmoudi}
\thanks{These authors contributed equally.}
\affiliation{Heyrovský Institute of the CAS, Dolejškova 2155/3, 182 23 Prague 8, Czech Republic}
\affiliation{School of Physical and Chemical Sciences, Queen Mary University of London, London, E1 4NS, United Kingdom} 

\author{Jiate Luo}
\thanks{These authors contributed equally.}
\affiliation{Department of Chemistry, Princeton University, New Jersey, USA}
\affiliation{Institute of Chemistry, Chinese Academy of Sciences, Beijing, China} 

\author{Chonghao Zhai}
\affiliation{State Key Laboratory for Mesoscopic Physics, School of Physics, Peking University, Beijing, China}

\author{Qiang Shi}
\affiliation{Institute of Chemistry, Chinese Academy of Sciences, Beijing, China} 

\author{Alexander Yosifov}
\affiliation{Clarendon Laboratory, University of Oxford, Parks Road, Oxford, OX1 3PU, United Kingdom}

\author{Jinzhao Sun}
\email{jinzhao.sun.phys@gmail.com}
\affiliation{School of Physical and Chemical Sciences, Queen Mary University of London, London, E1 4NS, United Kingdom}


\let\oldaddcontentsline\addcontentsline
\renewcommand{\addcontentsline}[3]{}

\begin{abstract}

Optical spectra of dissipative molecular systems depend on both complex resonance energies and the way each resonance couples to an optical probe. In a non-Hermitian description, these contributions are determined by complex residues involving left and right eigenvectors. We introduce a quantum protocol that reconstructs the resonance poles and their biorthogonal optical residues within a common spectral-filtering and measurement framework. A Hermitian dilation embeds the left and right eigenvectors into a zero-energy subspace. Gaussian-filtered correlators evaluated with unitary circuits, together with ratios of complex off-diagonal measurements, yield the residues without separate variational optimisation of the eigenstates. We demonstrate spectral reconstruction for dissipative plasmon-molecule models on IBM quantum hardware, resolving how probe-dependent modal contributions produce interference-induced doublets or accompany a separation of resonance energies. We also analyse the sensitivity of residue reconstruction near exceptional points. By making complex optical residues an explicit computational output, the protocol connects non-Hermitian eigenstructure to measurable optical response.
 

\end{abstract}

\maketitle

Interactions with the environment shape energy transport and photochemical processes in molecular materials and biological systems
\cite{tanimura2020numerically,PhysRevLett.120.030402,luo2026proton}. Dissipation and dephasing also complicate the connection between
measured optical spectra and the underlying resonances \cite{PhysRevLett.120.030402}.
When the relevant optical coherences admit an effective non-Hermitian description, complex eigenvalues encode resonance energies and decay rates \cite{Moiseyev_2011,Ashida02072020,El-Ganainy2018}. The contribution of each resonance to the optical response also
depends on a residue involving its left and right eigenvectors
\cite{FinkelsteinShapiro2023}. These contributions can interfere to
produce two spectral peaks even when the real pole positions coincide
\cite{Murata2015,FinkelsteinShapiro2023}.
Moreover, as an exceptional point (EP) is approached, eigenvalues and eigenvectors coalesce \cite{Miri2019,Naghiloo2019}, making the decomposition into individual resonance contributions ill-conditioned
\cite{exceptiona,Ashida02072020}. Connecting non-Hermitian modes to optical spectra therefore requires their complex poles and biorthogonal residues.


Quantum algorithms have expanded access to non-Hermitian spectral properties through variational eigenstate
preparation~\cite{Xie2024NonHermitianVQE}, eigenvalue estimation and
transformation~\cite{zhang2025exponential,Low2026EigenvalueProcessing},
pseudospectral analysis~\cite{Yang2026Pseudospectra}, and oblique
eigenprojection~\cite{Dalzell2026Oblique}. These approaches address complementary tasks. Eigenvalue and singular-value estimates identify spectral structure, but do not by themselves determine the complex residues of a chosen optical channel. Variational schemes can evaluate biorthogonal matrix elements using
separately optimised left and right
states~\cite{Xie2024NonHermitianVQE}, while quantum linear-algebra methods provide access to resolvent matrix
elements~\cite{Wang2024RandomizedLinearAlgebra}. Differences between expectation values evaluated in left and right eigenstates can also signal EPs \cite{Lakkaraju2026Observable}, illustrating how the relevant eigenstate information depends on the observable. In chemical spectroscopy of open molecular aggregates and lossy molecule-cavity systems
\cite{tanimura2020numerically,PhysRevLett.120.030402,FinkelsteinShapiro2023}, the task is to resolve each mode's contribution to a chosen optical channel, including modes with the same resonance frequency but different decay rates. Yet the real-frequency peak-ratio construction of
generalised quantum computational spectroscopy does not directly separate such modes \cite{zhai2025universal}. We show how a nonvariational algorithm built from unitary evolutions can extract both complex poles and the residues of a chosen optical channel.

Here we develop a quantum algorithm by embedding the shifted operator
$H_z=H-z\mathbb{I}$, for an effective non-Hermitian Hamiltonian $H$ and complex trial energy $z$, in a Hermitian dilation. At an eigenvalue, the dilation's zero-energy subspace contains separate embeddings of the corresponding left and right eigenvectors. Gaussian filtering targets this subspace, and randomised sampling of unitary evolutions estimates the matrix elements needed to
reconstruct the optical residues. We demonstrate the method numerically for molecular excitations coupled to
plasmonic cavity modes, an established setting for strong-coupling spectroscopy
\cite{Torma2015,Chikkaraddy2016,Hirai2023} and molecular polaritonics
\cite{10.1039/c8sc01043a,Molecular_Polaritonics,10.1021/acs.chemrev.2c00895,10.1021/acs.chemrev.2c00855,10.1021/acsphotonics.7b00680,Polariton_Chemistry}, including efforts to modify chemical dynamics
\cite{10.1021/acs.accounts.6b00295,PhysRevLett.116.238301,Galego2016} and energy transport \cite{Sandik2025} via strong coupling. For models with second- and third-order EPs, comparisons with exact diagonalisation validate the reconstructed complex poles and probe-dependent optical spectra. These results distinguish interference-induced doublets from genuine splitting of the resonance energies.

{\textit{Chemical non-Hermitian Hamiltonian.---}}We consider $N$ identical molecules uniformly coupled to one plasmon mode. The optically accessible single-excitation sector is spanned by $|P\rangle=a^\dagger|G\rangle$ and the collective molecular bright state $|X\rangle=N^{-1/2}\sum_n\sigma_n^+|G\rangle$, where $|G\rangle$ is the joint ground state. The resulting $2\times2$ model describes a collective bright-sector projection. The $N-1$ orthogonal molecular states are dark under the uniform-coupling and uniform-dipole assumptions.
In the basis $\{\vert P\rangle, \vert X\rangle\}$, the non-Hermitian Hamiltonian is 
\begin{equation}
H_{\mathrm{eff}}=
\begin{pmatrix}
\widetilde{\omega}_{P} & g\\
g & \widetilde{\omega}_{X}
\end{pmatrix},
\label{eq:two_mode_hamiltonian}
\end{equation}
where $\tilde{\omega}_P = \omega_{\mathrm{p}} - i\frac{\kappa}{2}$ and $\tilde{\omega}_X = \omega_{\mathrm{x}} - i\frac{\gamma}{2}$ denote the complex bare energies incorporating the transition frequencies $\omega_{\mathrm{p}}$ and $\omega_{\mathrm{x}}$ alongside their respective decay linewidths $\kappa$ and $\gamma$. Here $g$ is the collective plasmon-exciton coupling, and $\kappa/2$ and $\gamma/2$ are the total optical-coherence decay rates. As shown in SM Sec.~\ref{SM:sec6}, Lindblad recycling terms vanish on the optical-coherence block, making this effective Hamiltonian sufficient for linear absorption within the specified model. Diagonalisation of $H_{\mathrm{eff}}$ yields the exact complex eigenvalues 
\begin{equation}
z_\pm = \frac{\tilde{\omega}_P + \tilde{\omega}_X}{2} \pm \sqrt{\left(\frac{\tilde{\omega}_P - \tilde{\omega}_X}{2}\right)^2 + g^2}. 
\end{equation}
On resonance $\omega_{\mathrm{p}} = \omega_{\mathrm{x}} \equiv \bar{\omega}$, this simplifies to 
\begin{equation}
z_\pm = \bar{\omega} - i\frac{\kappa + \gamma}{4} \pm \sqrt{g^2 - \frac{(\kappa - \gamma)^2}{16}}.
\end{equation}
This reveals an EP at critical collective coupling $g_{\mathrm{EP}} = \frac{\vert{}\kappa - \gamma\vert{}}{4}$. Below this threshold (weak coupling), the bright poles
have the same real part but different linewidths; above it (strong coupling),
their real parts split while their linewidths coincide in the symmetric model. The EP is thereby controlled jointly by collective coherent coupling and dissipation, rather
than by the resonance of an isolated molecule with an
isolated plasmon. 

\textit{Hermitian dilation and biorthogonal eigenspaces.---}The chemical model shows that both complex poles and the associated left and right eigenvectors are needed. We now formulate a quantum algorithm for accessing them. Let $H\equiv H_{\mathrm{eff}}$, with right and left eigenvectors defined as $H\lvert r_\alpha\rangle=z_\alpha\lvert r_\alpha\rangle$ and $H^\dagger\lvert l_\alpha\rangle=z_\alpha^*\lvert l_\alpha\rangle$, respectively. For a complex trial energy $z$ and shifted operator $H_z$, the smallest-singular-value landscape is $f(z)\equiv\sigma_{\min}(H_z)$ \cite{Trefethen1999}. The condition $f(z)=0$ holds iff $z$ is an eigenvalue of $H$ (Proposition \ref{SM:prop1}). Moreover, under $H\rightarrow H+\Delta$, Weyl's inequality gives $\left|\sigma_{\min}(H+\Delta-z\mathbb{I})-\sigma_{\min}(H-z\mathbb{I})\right|\leq\|\Delta\|$, where $\| \cdot \|$ is the operator norm. Thus, finite-threshold regions of $f(z)$ provide a perturbation-controlled measure of near-singular spectral structure, including near an EP. The corresponding Lipschitz-continuity result and stability bounds for the detection regions are proved in Proposition \ref{SM:Lipschitz continuity} and Corollary \ref{SM:cor1}.

To access this landscape through unitary evolution, we introduce the Hermitian dilation 
$\mathcal{P}(z)=
\begin{pmatrix}
0 & H_z \\
H_z^\dagger & 0
\end{pmatrix},$ which acts on one additional two-level register and the original chemical Hilbert space. Its spectrum is $\operatorname{spec}\mathcal{P}(z)={\pm\sigma_k(H_z)}_k$, so locating an eigenvalue of $H$ is equivalent to locating a zero-energy eigenvalue of $\mathcal{P}(z)$ (Proposition \ref{SM:prop5}). The dilation also contains the eigenvectors. At $z=z_\alpha$, the embedded states $\binom{\lvert l_\alpha\rangle}{0}$ and $\binom{0}{\lvert r_\alpha\rangle}$ are zero modes of $\mathcal{P}(z_\alpha)$ (Proposition \ref{SM:prop6}). So for a simple eigenvalue, the corresponding kernel is $\operatorname{ker}\mathcal{P}(z_\alpha)=\operatorname{span}\!\left\{\binom{\lvert l_\alpha\rangle}{0},\binom{0}{\lvert r_\alpha\rangle}\right\}$. The dilation thus provides an orthogonal zero-energy subspace of a Hermitian operator containing both embedded members of the biorthogonal pair. This is the structure used below to reconstruct chemical observables. At the exact EP, the zero-mode construction remains well defined, but the simple-pole residue formula no longer applies; the response must be treated using a Jordan chain.

\textit{Gaussian filtering with unitary operations.---}The zero-energy subspace of $\mathcal{P}(z)$ can be isolated with a Gaussian spectral filter 
\begin{equation}
G_\tau(\mathcal{P}(z))
=
\exp\!\left[-\frac{\tau^{2}}{2}\mathcal{P}(z)^{2}\right],
\label{eq:gaussian_filter}
\end{equation}
where $\tau$ controls the spectral resolution (Lemma \ref{SM:lemma1} and Proposition \ref{SM:prop7}). Acting on an initial state $\lvert\Phi_0\rangle$ gives the filtered state $\lvert\Phi_\tau(z)\rangle=G_\tau(\mathcal{P}(z))\lvert\Phi_0\rangle$. Since $\mathcal{P}(z)$ is Hermitian, the filter suppresses components with nonzero dilation energies and preserves the zero-energy subspace. For an isolated zero-energy subspace, the filtering error decreases as $\exp[-\tau^2\Delta_\sigma^2/2]$, where $\Delta_\sigma$ is the smallest nonzero singular value of $H_z$. Eq. \ref{eq:gaussian_filter} can be implemented using only unitary operations. Its Fourier representation is 
\begin{equation}
G_\tau(\mathcal{P}(z))
=
\int_{-\infty}^{\infty}
\frac{dx}{\sqrt{2\pi}}\,
e^{-x^{2}/2}e^{-i\tau x\mathcal{P}(z)},
\end{equation}
where a finite quadrature gives $G_\tau(\mathcal{P}(z))\simeq\sum_m w_m e^{-i\tau x_m\mathcal{P}(z)}$ (proof in SM Sec. \ref{sec:spectral_filter_lcu}). Hence, randomised sampling of the quadrature points estimates observables associated with the discretised Gaussian filter through repeated unitary experiments, without non-unitary evolution under $H_z$ (Proposition \ref{SM:prop11}).

\begin{figure}[t]
    \centering
    \includegraphics[width=0.48\textwidth]{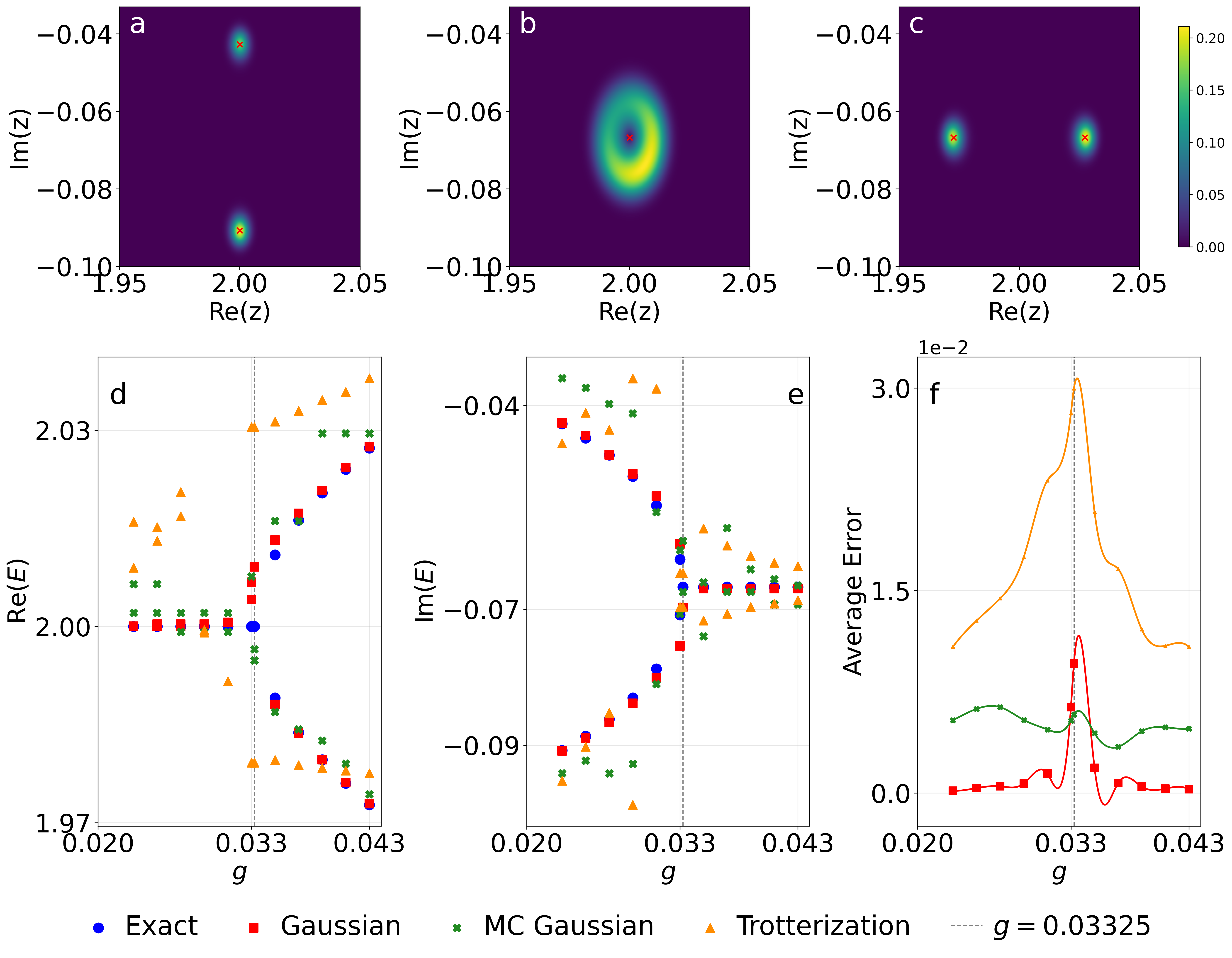}        
    \caption{Validation of the Gaussian-filtering algorithm around the EP. (a)--(c) Filtered signal $S_\tau(z)$ in the complex plane ($\tau = 40$, 250 samples) across three regimes: (a) weak coupling, (b) the EP, and (c) strong coupling, with exact eigenvalues marked by red crosses. (d) $\mathrm{Re}(E)$ and (e) $\mathrm{Im}(E)$ energy spectra versus $g$, highlighting Rabi splitting in the strong-coupling regime. Exact eigenvalues (blue) are compared against Gaussian (red squares), MC Gaussian (green crosses), and Trotterisation (orange triangles) reconstructions. (f) Average peak distance error versus $g$, calculated over 15 random initial states.}
    \label{fig:spectr}
\end{figure}

To locate the complex poles, we scan $z$ over the complex plane and evaluate the filtered signal $S_\tau(z)=\langle\Phi_\tau(z)\vert\Phi_\tau(z)\rangle$. The signal is enhanced when $H_z$ approaches singularity, and we use the maxima of $S_\tau(z)$ as numerical estimates of the complex eigenvalues. In practice, several initial states can be used to ensure adequate overlap with each zero-energy sector [Fig. \ref{fig:spectr}(a-c)]. The filter isolates the left-right eigenspace; the embedded measurements we introduce below then separate the two components and convert the filtered subspace into biorthogonal observables.

\textit{Biorthogonal reconstruction of optical response.---}Locating the complex poles is only the first step: their contributions
to an optical channel are determined by the associated residues. Away from the EP, where $H$ is diagonalisable and $\langle l_\alpha|r_\alpha\rangle\neq0$, its resolvent has the expansion $(\omega-H)^{-1}=\sum_\alpha \lvert r_\alpha\rangle\langle l_\alpha\rvert/[(\omega-z_\alpha)\langle l_\alpha\vert r_\alpha\rangle]$. For a preparation state $\lvert s\rangle$ and a detection state $\langle d\vert$, the residue associated with $z_\alpha$ is hence
\begin{equation}
f_\alpha=\frac{\langle d\vert r_\alpha\rangle\langle l_\alpha\vert s\rangle}{\langle l_\alpha\vert r_\alpha\rangle},
\label{eq:f_alpha}
\end{equation}
where $f_\alpha$ is generally complex and depends on the mode and the probe channel. For linear absorption, $\lvert s\rangle$ and $\langle d\vert$ are generated by the transition dipole, with $\lvert s\rangle=\hat{\mu}\lvert 0\rangle$ and $\langle d\vert=\langle 0\vert\hat{\mu}^\dagger$. The response is then reconstructed from the filtered poles and residues as $A(\omega)\propto-\operatorname{Im}\sum_\alpha f_\alpha/(\omega-z_\alpha)$ (SM Sec. \ref{SM:sec6}). This pole-residue form makes the origin of interference explicit: two modes can produce a split or asymmetric line shape even when their real pole positions are not separated. For the matched dipole channel $\langle d|=\langle s|$, the optical residue generally differs from the right-state weight
$|\langle r_\alpha|s\rangle|^2/\langle r_\alpha|r_\alpha\rangle$. The latter is an ordinary expectation value in a prepared right
eigenstate, whereas the residue also contains the dual-basis excitation coefficient. These quantities coincide for an orthonormal eigenbasis. Although the complete right-eigenvector matrix determines the dual
basis through its inverse, the dilation gives direct access to the required left-right matrix elements. 

For a chemical operator $O$, suppose the off-diagonal embedded operators $\mathcal{O}=\begin{pmatrix}0&O\\0&0\end{pmatrix}$ and $\mathcal{I}=\begin{pmatrix}0&\mathbb{I}\\0&0\end{pmatrix}$, where $\mathcal I$ is the off-diagonal embedding of the chemical identity. If the filtered state has nonzero overlap with both embedded components, their common preparation amplitudes cancel in the ratio
\begin{equation}
\frac{\langle\mathcal{O}\rangle}{\langle\mathcal{I}\rangle}=\frac{\langle l_\alpha\vert O\vert r_\alpha\rangle}{\langle l_\alpha\vert r_\alpha\rangle}.
\label{eq:ratio}
\end{equation}
To make that explicit, consider the zero-energy component of the filtered state at $z=z_\alpha$, written as $|\Psi_{\alpha,0}\rangle=a_\alpha\bigl(|l_\alpha\rangle,0\bigr)^{\mathsf T}+b_\alpha\bigl(0,|r_\alpha\rangle\bigr)^{\mathsf T},$ with $a_\alpha^{*}b_\alpha\neq0$. For the embedded operators $\mathcal{O}$ and $\mathcal{I}$, direct block multiplication gives $\langle\Psi_{\alpha,0}|\mathcal{O}|\Psi_{\alpha,0}\rangle=a_\alpha^{*}b_\alpha\langle l_\alpha|O|r_\alpha\rangle$ and $\langle\Psi_{\alpha,0}|\mathcal{I}|\Psi_{\alpha,0}\rangle=a_\alpha^{*}b_\alpha\langle l_\alpha|r_\alpha\rangle$. Their ratio hence gives Eq. \ref{eq:ratio}. Choosing $O=|s\rangle\langle d|$ we get Eq. \ref{eq:f_alpha}, the pole residue entering the optical response. The off-diagonal operators are decomposed into Hermitian components using the dilation-register Pauli operators, so their real and imaginary parts can be measured with standard protocols. This converts the filtered zero-energy subspace into the biorthogonal observables required for non-Hermitian spectroscopy. Direct Gaussian filtering of a non-Hermitian operator is not an orthogonal spectral projection.
The Fourier representation of the direct Gaussian filter is generally not realisable as a RLCU constructed from unitary evolutions.


\begin{figure}[t]
    \centering
    \includegraphics[width=0.48\textwidth]{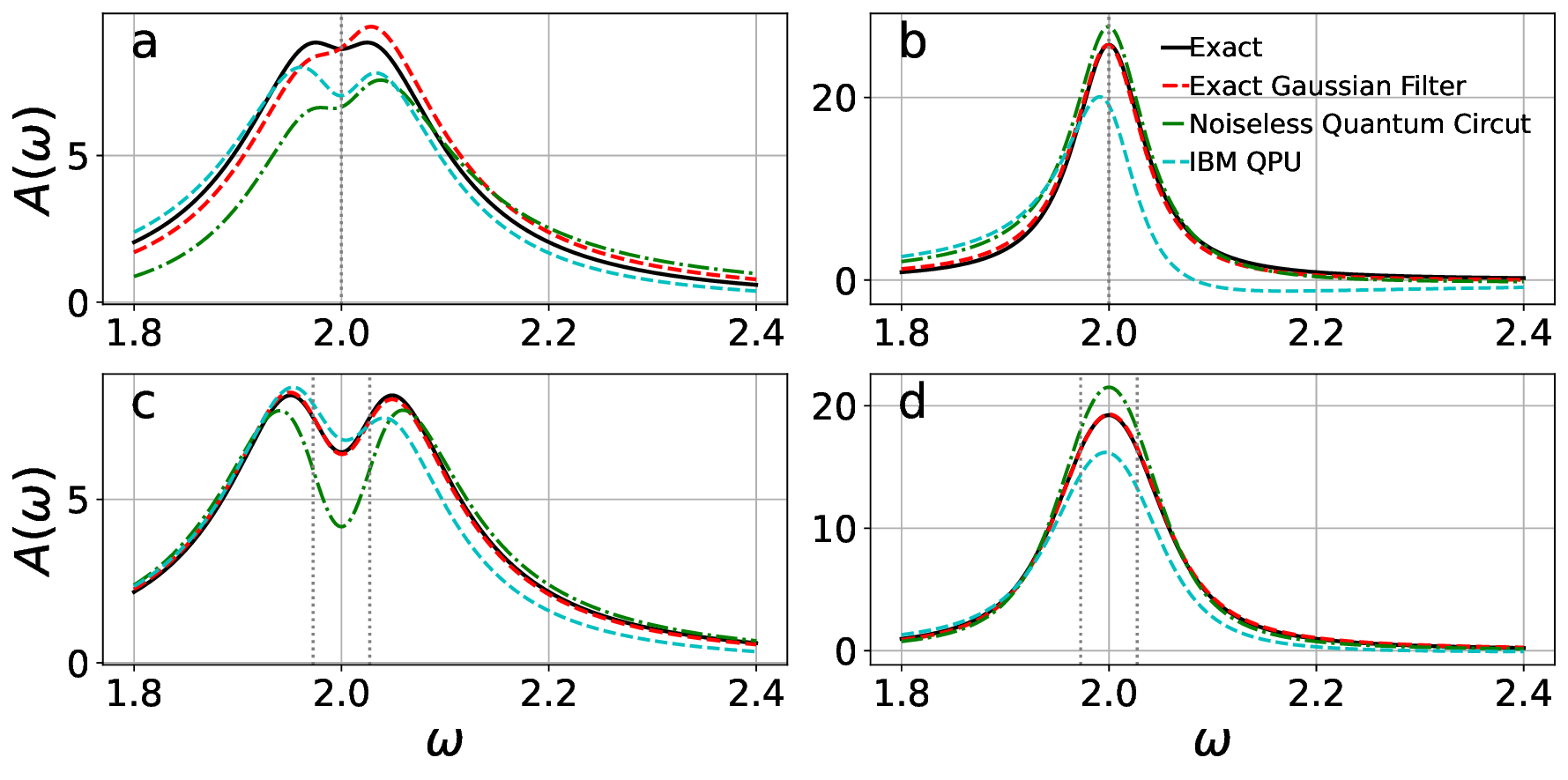}        
     
    \caption{Reconstructed effective absorption spectra $A(\omega)$ under distinct coupling regimes and dipole configurations. (a),(b) Weak-coupling regime $g = 0.023$ for plasmon-dominant ($\mu_P=1, \mu_X=0$) and exciton-dominant ($\mu_P=0, \mu_X=1$) dipoles. (c),(d) Strong-coupling regime $g = 0.043$ displaying Rabi splitting for plasmon-dominant dipole. Spectra obtained via exact Hamiltonian diagonalisation are compared against exact Gaussian filtering, noiseless circuit MC simulation ($\tau=35$, $30$ samples), and IBM quantum hardware results. Vertical dotted lines denote the real parts of the complex resonance energies.}
    \label{fig:response}
\end{figure}

\textit{Interference splitting versus Rabi splitting.---}We now test the complete procedure by varying the molecular-plasmon coupling $g$ through the EP of the plasmon-molecular excitation model. For each value of $g$, we obtain reference poles by direct diagonalisation and independently scan $z$ using the filtered signal $S_\tau(z)$. The proposed pole estimates are obtained from the maxima of $S_\tau(z)$. These estimates reproduce both the real and imaginary parts of the exact
poles [Fig. \ref{fig:spectr}(d,e)], with average peak-distance errors shown in [Fig. \ref{fig:spectr}(f)].

The pole trajectories establish the distinction between two spectroscopically different regimes. Below the EP, the two bright poles have the same real but different imaginary parts. Their different linewidths and complex residues can nevertheless interfere to produce two apparent maxima in the optical response. The resulting doublet is hence an interference feature rather than evidence of two distinct resonance energies. Above the EP, the real parts of the poles split by $2\sqrt{g^2 - g_{\mathrm{EP}}^2}$ while their linewidths become equal in the symmetric model. In this regime, the optical splitting is accompanied by eigenvalue separation and corresponds to Rabi-like splitting.

\begin{figure}[t]
    \centering
    \includegraphics[width=0.99\linewidth]{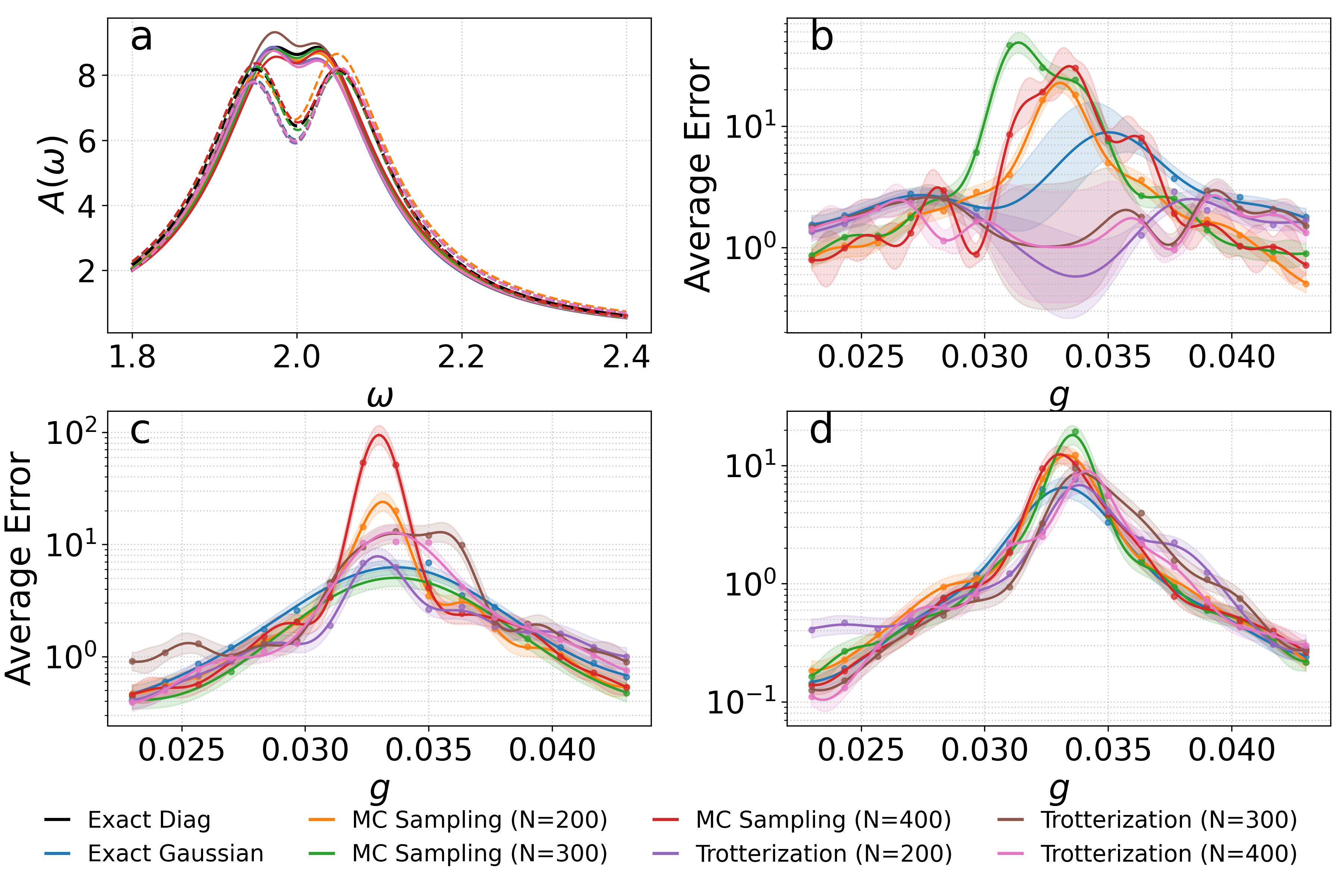}
    \caption{Optical spectra and average reconstruction errors across coupling strength $g$ averaged over 15 initial states. (a) Effective optical response $A(\omega)$ at $g = 0.023$ (solid) and $g = 0.043$ (dashed). (b)–(d) Average reconstruction error as a function of $g$ for filter times (b) $\tau = 40$, (c) $\tau = 80$, and (d) $\tau = 160$ using Gaussian process regression fits with $95\%$ confidence bands, showing sharp error peaks near the EP.}
    \label{fig:combined_spectra_and_error}
\end{figure}

The response also depends on how the system is prepared and measured. Mixed, molecular- and plasmon-only probes produce different residues for the same pole structure. Fig. \ref{fig:response} compares the reconstructed responses for different dipole
channels, while Fig. \ref{fig:combined_spectra_and_error} summarizes the coupling dependence of the spectra and reconstruction errors. This channel dependence shows why eigenvalue detection alone cannot reproduce non-Hermitian spectroscopy. The combined pole and residue analysis provides a stricter interpretation of spectral splitting in dissipative molecular systems. A doublet should be identified as genuine Rabi splitting only when it is accompanied by separated real poles in the complex spectrum; otherwise, it may result from interference between modes with different decay rates. The Gaussian-filtering procedure supplies both pieces of information within a unitary-compatible framework.

At the EP, it is the decomposition into separate modes that becomes singular, not the damped optical response. As the two poles coalesce, individual residues can diverge while their contributions cancel in the total polarizability. Writing $H=z_{\mathrm{EP}}\mathbb I+J$, with $J^2=0$ on the two-dimensional bright sector gives
\begin{equation}
\alpha_{+,\mathrm{EP}}(\omega)
=-\frac{\langle s|s\rangle}{\omega-z_{\mathrm{EP}}}
-\frac{\langle s|J|s\rangle}{(\omega-z_{\mathrm{EP}})^2},
\label{eq:main_ep_response}
\end{equation}
where $z_{\mathrm{EP}}=\bar\omega-i(\kappa+\gamma)/4$ in the linewidth convention. This expression is finite at real probe frequencies for positive damping and is equivalent to SM Eq.~\ref{eq:alpha_EP}, where the rates denote coherence decay rates. The dipole-dependent coefficients show why an EP fixes neither the number of spectral maxima nor a universal squared-Lorentzian line shape. The cancellation also explains why reconstruction becomes sensitive near the EP, as reflected in Fig.~\ref{fig:combined_spectra_and_error}. The dilation kernel remains well defined there, but ordinary zero-mode filtering alone does not supply the higher-order resolvent coefficient.

\begin{figure}[t]
    \centering    \includegraphics[width=0.48\textwidth]{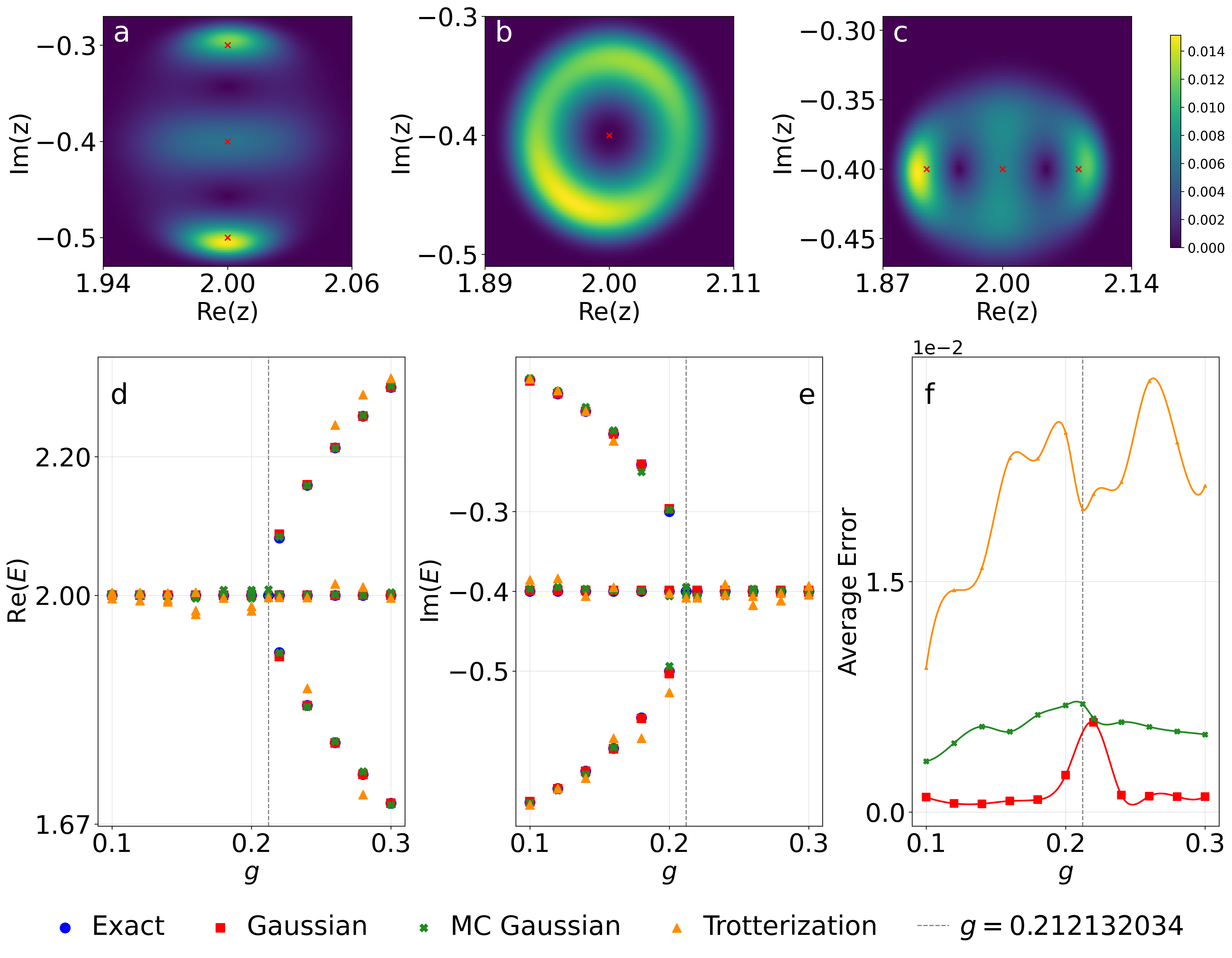}        
    \caption{Higher-order validation of the Gaussian-filtering algorithm around the EP3. Extension of Fig. \ref{fig:spectr} to Eq. \ref{eq:three_mode}. Using the same filtering and reconstruction parameters, (a)--(c) show $S_{\tau}(z)$, respectively, in the weak-coupling, EP3, and strong-coupling regimes. (d),(e) The real and imaginary pole trajectories, including three-branch coalescence and bifurcation. (f) The average peak-distance error over 15 initial states.}
    \label{fig:spectr_EP3}
\end{figure}

\textit{Beyond the two-mode model.---}The construction is not restricted to a single bright molecular mode or a second-order EP. As a higher-order example [Fig. \ref{fig:spectr_EP3}], consider two plasmon modes with no direct intermode coupling and one collective molecular excitation (i.e., EP3) \cite{Laha2020}. In the basis
$\{\lvert P_{1}\rangle,\lvert X\rangle,\lvert P_{2}\rangle\}$, let
\begin{equation}
H_{3}=
\begin{pmatrix}
\omega_{0}-i(\Gamma_{0}-\eta) & g & 0\\
g & \omega_{0}+\delta-i\Gamma_{0} & g\\
0 & g & \omega_{0}-i(\Gamma_{0}+\eta)
\end{pmatrix},
\label{eq:three_mode}
\end{equation}
where $g=\sqrt{N}g_{0}$ and $\Gamma_{0}>\eta>0$. Here $\delta$ is the molecular detuning and $\eta$ is the linewidth imbalance between the two plasmon modes. Subtracting the common complex energy $\omega_{0}-i\Gamma_{0}$ gives the shifted core $
H_{3}-(\omega_{0}-i\Gamma_{0})\mathbb{I}_{3} =
\begin{pmatrix}
i\eta & g & 0\\
g & \delta & g\\
0 & g & -i\eta
\end{pmatrix},$
which is $\mathcal{PT}$-symmetric for real $\delta$, while the uniform shift ensures the physical system remains passive.

\begin{figure}[t]
    \centering    \includegraphics[width=0.48\textwidth]{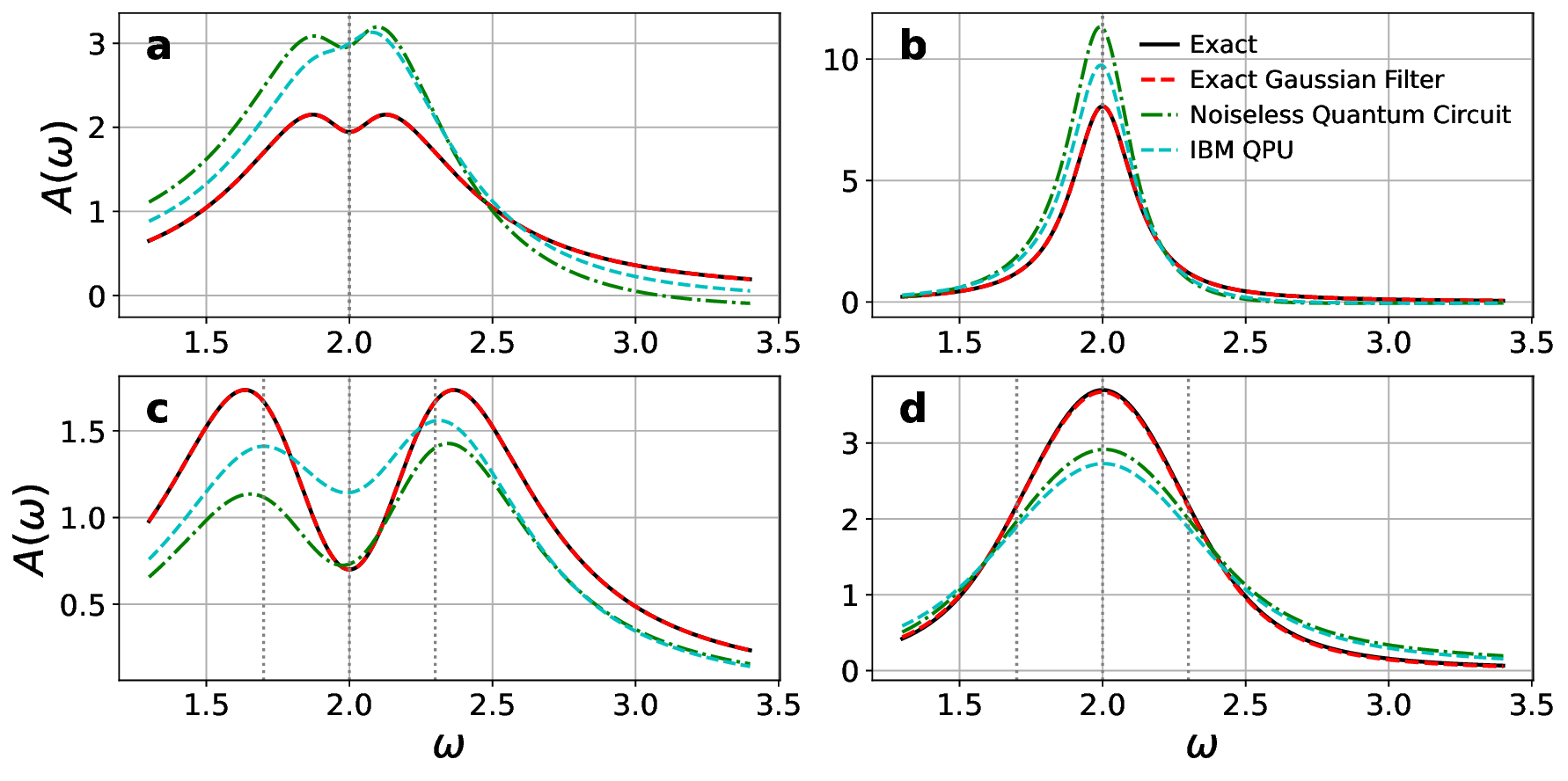}        
   \caption{Extension of Fig. \ref{fig:response} to EP3. (a),(b) Weak-coupling spectra with interference-induced doublets. (c),(d) Strong-coupling spectra with genuine pole splitting and channel-dependent line shapes. Vertical dotted lines mark the real parts of the complex poles.}
\label{fig:ep3_spectra}
    \label{fig:rabi_splitting_EP3}
\end{figure}

At resonance $\delta=0$, the poles are
$z_{0}=\omega_{0}-i\Gamma_{0}$ and
$z_{\pm}=\omega_{0}-i\Gamma_{0}\pm\sqrt{2g^{2}-\eta^{2}}$. Consequently, EP3 occurs at
$g_{\mathrm{EP3}}=\eta/\sqrt{2}$, where all three poles and their right eigenvectors coalesce at
$z_{\mathrm{EP3}}=\omega_{0}-i\Gamma_{0}$. The coalesced right eigenvector is proportional to
$\lvert P_{1}\rangle-i\sqrt{2}\lvert X\rangle-\lvert P_{2}\rangle$. For $g<g_{\mathrm{EP3}}$, the three poles have the same real part but distinct decay rates; for $g>g_{\mathrm{EP3}}$, the pair $z_{\pm}$ separates in real energy. At the EP3, the resolvent can contain poles of order up to three \cite{Hodaei2017}. Away from EP3, the Hermitian-dilation, filtering, and residue-reconstruction procedures are unchanged.
As shown in Fig. \ref{fig:spectr_EP3}, the Gaussian-filtering algorithm remains numerically stable across the EP3, accurately recovering the complex spectrum without singular divergence near the coalescence point. Furthermore, the reconstructed response functions show that interference-induced line-shape splitting persists in the weak-coupling regime $g < g_{\mathrm{EP3}}$ prior to genuine real-energy pole separation, Fig. \ref{fig:rabi_splitting_EP3}.

\textit{Discussion.---}Spectroscopy in dissipative molecular systems requires distinguishing changes in resonance energies from changes in how modes contribute
to an optical channel. Our protocol makes both quantities accessible through unitary quantum experiments by estimating complex poles and
biorthogonal residues. The plasmon-molecule examples show why this matters: overlapping modes can produce a doublet without real-energy
pole splitting, and changing the probe can reshape the spectrum without changing the poles. The reconstructed response thus connects
non-Hermitian mode structure to observable line shapes. The second- and third-order EP examples show that this distinction also applies at higher-order coalescence. At an EP,
the total damped response remains finite even as the individual residues are ill-conditioned. Operationally, the Hermitian dilation turns the non-Hermitian spectral problem into filtering and measurement with unitary evolutions. The filtered zero modes encode the left-right matrix
elements needed to calculate optical residues, which combine with the poles to reconstruct channel-resolved spectra. Our circuit and calculations on IBM hardware verify this. Applying the protocol to realistic aggregates can support model-based spectral interpretation and probe design.

\textit{Acknowledgments.---}P.M. is supported by the Grant Agency of the Czech Republic (Grant No. 24-10992S). A.Y. is supported by the UKRI Future Leaders Fellowship (Grant No. 10128920). We acknowledge the use of IBM Quantum services for this work. The views expressed are those of the authors and do not reflect the official policy or position of IBM or the IBM Quantum team. 

\textit{Data availability.---}The data are available upon reasonable request from the authors.


\bibliography{Ref}

\clearpage
\widetext

\begin{center}
{\large\bfseries Supplemental Material for\\[0.35em]
``Quantum computational spectroscopy of complex resonances and optical residues''}
\end{center}
\label{sec:SM}

\section{Mathematical Formulation of the Non-Hermitian Spectral Problem}
\label{SM:sec1}
The spectrum of a non-Hermitian operator is generally not a stable object:
arbitrarily small perturbations may induce large spectral displacements for
non-normal operators. A perturbation-stable characterisation is instead
provided by the pseudospectrum, which can be characterised through the smallest singular value of the shifted operator $H_z=H-z\mathbb{I}$. Specifically, the zeros of the smallest singular value recover the spectrum,
while its sublevel sets define the pseudospectrum. Consequently, the non-Hermitian spectral problem can be reformulated as the construction of
the singular-value landscape over the complex plane. This reformulation provides the mathematical foundation of the quantum algorithm developed in
this work.

\subsection{Spectrum and Pseudospectrum}

The eigenvalues of a Hermitian operator are stable under small
perturbations. In contrast, the spectrum of a non-Hermitian operator may
change dramatically under arbitrarily small perturbations, particularly
for non-normal matrices. Consequently, the spectrum alone does not
provide a robust characterisation of the spectral properties of a
non-Hermitian operator. To account for this sensitivity, one introduces the
\(\varepsilon\)-pseudospectrum, defined as
\begin{equation}
\Lambda_\varepsilon (H)
=
\bigcup_{\|\Delta\|\le\varepsilon}
\operatorname{spec}(H+\Delta),
\label{eq:pseudo_def}
\end{equation}
where the union is taken over all perturbations satisfying
\(\|\Delta\|\le\varepsilon\).
Thus, \(\Lambda_\varepsilon(H)\) consists of all points in the complex
plane that become eigenvalues after perturbations of size at most
\(\varepsilon\). The spectrum is recovered as the limiting case $\Lambda_0(H)=\operatorname{spec}(H).$
Although Eq.~\ref{eq:pseudo_def} provides a robust mathematical
definition, it is not directly suitable for computation since it
requires considering infinitely many perturbations. The following result
shows that the entire pseudospectrum can instead be characterised by a
single scalar quantity.

\begin{proposition}[Singular-value characterisation]
For every \(z\in\mathbb C\)
\begin{equation}
z\in\Lambda_\varepsilon(H)
\iff
\sigma_{\min}(H_z)
\le
\varepsilon,
\label{eq:sv_pseudo}
\end{equation}
where \(\sigma_{\min}(\cdot)\) denotes the smallest singular value. Consequently
\begin{equation}
z\in\operatorname{spec}(H)
\iff
\sigma_{\min}(H_z)=0.
\label{eq:sv_spectrum}
\end{equation}
\label{SM:prop1}
\end{proposition}

\begin{proof}
For any matrix \(A\), the Eckart-Young theorem states that the distance
from \(A\) to the nearest singular matrix equals its smallest singular value $\sigma_{\min}(A)=\min_{\det(A+\Delta)=0}
\|\Delta\|.$ Applying this identity to $H_z$ gives $\sigma_{\min}(H_z) = \min_{\det(H_z+\Delta)=0} \|\Delta\|.$
Therefore, $\sigma_{\min}(H_z)\le\varepsilon$
iff there exists a perturbation satisfying
\(\|\Delta\|\le\varepsilon\) such that
\(
\det(H_z+\Delta)=0,
\) which is precisely the definition of
\(z\in\Lambda_\varepsilon(H)\).
Setting \(\varepsilon=0\) immediately yields
Eq.~\ref{eq:sv_spectrum}.
\end{proof}

Proposition \ref{SM:prop1} shows that both the spectrum and the pseudospectrum are
completely determined by the smallest singular value of $H_z$. Consequently, the non-Hermitian eigenvalue problem can be reformulated as the construction of the scalar function $f_H(z)=\sigma_{\min}(H_z)$ over the complex plane. The eigenvalues of \(H\) correspond to the zeros
of \(f_H\), while its sublevel sets determine the pseudospectrum.
Therefore, rather than searching directly for the roots of
\(\det(H_z)\), the remainder of this work is devoted to constructing and measuring the singular-value landscape \(f_H(z)\).

\subsection{Stability of the Singular-Value Landscape}

For the landscape $f_H(z)$ to serve as the basis of a robust quantum algorithm, it must vary continuously under perturbations of the underlying non-Hermitian operator. The following establishes this stability.

\begin{proposition}[Lipschitz continuity]
\label{SM:Lipschitz continuity}
Let \(
H'
=
H+\Delta,
\) where \(
\|\Delta\|
\le
\delta.
\) Then for every \(z\in\mathbb C\)
\begin{equation}
|f_{H'}(z)-f_H(z)|
\le
\delta.
\label{eq:lipschitz}
\end{equation}
\end{proposition}

\begin{proof}

By definition, \(f_H(z)
=
\sigma_{\min}(H_z)
\)
and
\(
f_{H'}(z)
=
\sigma_{\min}(H_z+\Delta).
\)
Weyl's perturbation theorem for singular values states that $|\sigma_j(A)-\sigma_j(B)| \le \|A-B\|$ for every singular value. Choosing \(
A
=
H_z\) and \(B
=
H_z+\Delta
\) gives
\(
A-B
=
-\Delta.
\)
Therefore, $|f_{H'}(z)-f_H(z)| = |\sigma_{\min}(A)-\sigma_{\min}(B)| \le
\|\Delta\|,$ which proves Eq.~\ref{eq:lipschitz}.
\end{proof}

Proposition \ref{SM:Lipschitz continuity} shows that the singular-value landscape depends continuously on perturbations of the underlying operator. Consequently, small perturbations cannot produce large pointwise
distortions of \(f_H(z)\). In particular, if
\(
z^\ast
\)
is an isolated zero of \(f_H\), then every sufficiently small perturbation produces a nearby minimum of the perturbed landscape. Hence, the localisation of eigenvalues can be formulated as the
problem of tracking the minima of a stable scalar function over the complex plane. Unlike the eigenvalues themselves, which may exhibit arbitrarily large displacements near defective points, the singular-value landscape remains uniformly Lipschitz continuous with respect to operator perturbations. As a result, the quantum algorithm constructs a perturbation-stable object even when the spectrum is highly ill-conditioned.

\begin{corollary}[Stability of detection regions]
For $\epsilon\ge0$, define the detection region
\[
\mathcal Z_\epsilon(H)
=
\left\{
z\in\mathbb C:
f_H(z)\le\epsilon
\right\},
\]
where
\(
f_H(z)=\sigma_{\min}(H_z)
\).
If
\(
\|\Delta\|\le\delta,
\)
then $\mathcal Z_\epsilon(H) \subseteq
\mathcal Z_{\epsilon+\delta}(H+\Delta)$ and $\mathcal Z_\epsilon(H+\Delta) \subseteq
\mathcal Z_{\epsilon+\delta}(H).$
\label{SM:cor1}
\end{corollary}

\begin{proof}
Let
\(
z\in\mathcal Z_\epsilon(H)
\).
Then 
\(
f_H(z)\le\epsilon.
\) By Proposition~\ref{SM:Lipschitz continuity}
\[
f_{H+\Delta}(z)
\le
f_H(z)+\delta
\le
\epsilon+\delta,
\]
which implies $z\in \mathcal Z_{\epsilon+\delta}(H+\Delta).$ Hence, $\mathcal Z_\epsilon(H) \subseteq
\mathcal Z_{\epsilon+\delta}(H+\Delta).$ Interchanging the roles of \(H\) and \(H+\Delta\) yields $\mathcal Z_\epsilon(H+\Delta) \subseteq \mathcal Z_{\epsilon+\delta}(H),$ which completes the proof.
\end{proof}
Hence, perturbations of size at most $\delta$ modify the near-zero regions of the singular-value landscape only through an additive increase of the detection threshold. Although the eigenvalues
of a non-normal operator may undergo large displacements under small
perturbations, the singular-value landscape itself remains uniformly
stable. Consequently, the quantum algorithm constructs a perturbation-stable detection landscape whose near-zero regions provide a robust characterisation of the spectrum and the pseudospectrum. Together, Proposition \ref{SM:prop1}, Proposition \ref{SM:Lipschitz continuity}, and Corollary \ref{SM:cor1} establish that the
non-Hermitian spectral problem can be reformulated as the construction
of a perturbation-stable singular-value landscape. The remaining task is to develop an efficient quantum procedure for evaluating
\(f_H(z)=\sigma_{\min}(H_z)\).

\subsection{Hermitian Reformulation of the Singular-Value Problem}
The remaining challenge is that singular values are not directly accessible through unitary quantum evolution. Our strategy is thus to embed the shifted operator $H_z$ into a Hermitian operator whose
energy spectrum reproduces its singular values. Once this is achieved, the smallest singular value can be obtained by detecting the lowest energy of a Hermitian Hamiltonian. For each complex parameter \(z\), define $A(z)=H_z-zI,$ and introduce the Hermitian dilation
\begin{equation} \nonumber
\mathcal{P}(z)=
\begin{pmatrix}
0&A(z)\\
A(z)^\dagger&0
\end{pmatrix},
\label{eq:dilation}
\end{equation}
which acts on the enlarged Hilbert space $\mathcal{H}\oplus \mathcal{H}$. Since \(\mathcal{P}(z)=\mathcal{P}(z)^\dagger\), the family $U(t)=\exp{(-it\mathcal{P}(z))}$ is unitary. Consequently, the non-Hermitian spectral problem has been converted into a Hamiltonian-simulation problem for a Hermitian operator. The key property of this construction is that the spectrum of $\mathcal{P}(z)$ is determined
entirely by the singular values of \(A(z)\).

\begin{proposition}[Hermitian dilation]
\label{SM:prop5}
For every \(z\in\mathbb C\)
\begin{equation}
\operatorname{spec}(\mathcal{P}(z))
=
\{\pm\sigma_k(H_z)\}_k .
\label{eq:dilation_spectrum}
\end{equation}
As a result
\begin{equation}
\min_{\lambda\in\operatorname{spec}(\mathcal{P}(z))}
|\lambda|
=
\sigma_{\min}(H_z),
\label{eq:min_sv}
\end{equation}
and thus
\begin{equation}
z\in\operatorname{spec}(H)
\iff
\mathcal{P}(z) \
\text{possesses a zero-energy eigenstate.}
\label{eq:zero_mode}
\end{equation}
\label{SM:prop3}
\end{proposition}

\begin{proof}
Let $
\mathcal{P}(z)
\begin{pmatrix}
u\\
v
\end{pmatrix}
=
\lambda
\begin{pmatrix}
u\\
v
\end{pmatrix},$ where \(\lambda\) is an eigenvalue of $\mathcal{P}(z)$. From the definition of $\mathcal{P}(z)$, the eigenvalue equation is equivalent to the coupled system $
A(z)v=\lambda u,$ $A(z)^\dagger u=\lambda v.$ Applying \(A(z)^\dagger\) to the first equation and substituting the second yields $A(z)^\dagger A(z)v = \lambda A(z)^\dagger u = \lambda^2 v.$ Thus, \(\lambda^2\) is an eigenvalue of
\(A(z)^\dagger A(z)\). By definition, the singular values of \(A(z)\)
are the positive square roots of the eigenvalues of
\(A(z)^\dagger A(z)\). Hence, $\lambda^2=\sigma_k(A(z))^2,$ which implies $
\lambda=\pm\sigma_k(A(z)).$ This proves Eq.~\ref{eq:dilation_spectrum}. Eq.~\ref{eq:min_sv} follows immediately, since the eigenvalues of
$\mathcal{P}(z)$ occur in symmetric pairs
\(\pm\sigma_k(A(z))\), and so the smallest absolute eigenvalue equals
\(\sigma_{\min}(A(z))=\sigma_{\min}(H_z)\). Finally
\[
\sigma_{\min}(H_z)=0
\iff
H_z \
\text{is singular}
\iff
z\in\operatorname{spec}(H),
\] which is equivalent to the existence of a zero-energy eigenstate of
$\mathcal{P}(z)$ by Eq.~\ref{eq:min_sv}. This establishes
Eq.~\ref{eq:zero_mode}.
\end{proof}
Proposition \ref{SM:prop3} converts the original non-Hermitian eigenvalue problem
into the detection of zero-energy states of the Hermitian family $\mathcal{P}(z)$. Constructing the singular-value landscape is thereby equivalent to measuring the lowest-energy branch of a Hermitian Hamiltonian, enabling the use of unitary quantum simulation techniques.

\subsection{Recovery of the Biorthogonal Eigenspace}
\label{sec:SM1D}

Locating the eigenvalues alone, however, is insufficient for evaluating physical observables. Unlike the Hermitian case, expectation values are determined by the biorthogonal expression
\begin{equation}
\langle O\rangle_j
=
\frac{\langle L_j|O|R_j\rangle}
{\langle L_j|R_j\rangle},
\label{eq:bio_expectation}
\end{equation}
which depends simultaneously on the left and right eigenvectors. The
following proposition shows that these vectors are naturally embedded in
the zero-energy subspace of the Hermitian dilation.

\begin{proposition}[Zero modes and biorthogonal eigenvectors]
\label{SM:prop6}

Let \(\lambda_j\) be an eigenvalue of \(H\), and define $A(\lambda_j)=H-\lambda_j \mathbb{I}.$
Then the embedded states
\begin{equation}
|\chi_j^{\,L}\rangle
=
\begin{pmatrix}
|L_j\rangle\\
0
\end{pmatrix},
\qquad
|\chi_j^{\,R}\rangle
=
\begin{pmatrix}
0\\
|R_j\rangle
\end{pmatrix},
\end{equation}
span the kernel of the Hermitian dilation
\begin{equation}
\ker \mathcal{P}(\lambda_j)
=
\operatorname{span}
\left\{
|\chi_j^{\,L}\rangle,
|\chi_j^{\,R}\rangle
\right\}.
\end{equation}
\end{proposition}

\begin{proof}
Since $
H|R_j\rangle=\lambda_j|R_j\rangle$ and $\langle L_j|H=\lambda_j\langle L_j|,$ the shifted operator satisfies $A(\lambda_j)|R_j\rangle=0,$ $A(\lambda_j)^\dagger|L_j\rangle=0.$ Substituting these relations into the definition of the Hermitian
dilation gives $\mathcal{P}(\lambda_j) |\chi_j^{\,R}\rangle =
\begin{pmatrix}
A(\lambda_j)|R_j\rangle\\
0
\end{pmatrix}
=
0,$
and $\mathcal{P}(\lambda_j) |\chi_j^{\,L}\rangle =
\begin{pmatrix}
0\\
A(\lambda_j)^\dagger|L_j\rangle
\end{pmatrix}
=
0,$
establishing that both embedded vectors belong to the zero-energy subspace. Conversely, every zero mode of $\mathcal{P}(\lambda_j)$ satisfies $A(\lambda_j)|v\rangle=0$ and $A(\lambda_j)^\dagger|u\rangle=0,$ and therefore is a linear combination of the embedded right and left
eigenvectors. Hence $\ker \mathcal{P}(\lambda_j) =
\operatorname{span}
\left\{
|\chi_j^{\,L}\rangle,
|\chi_j^{\,R}\rangle
\right\}.$
\end{proof}
Proposition \ref{SM:prop6} establishes that the Hermitian dilation preserves not
only the spectral information encoded by the singular-value landscape,
but also the complete biorthogonal eigenspace of the original
non-Hermitian operator. Consequently, isolating the zero-energy subspace
of $\mathcal{P}(\lambda_j)$ simultaneously identifies the target eigenvalue and
provides direct access to the left and right eigenvectors required for
evaluating Eq.~\ref{eq:bio_expectation}. So the Hermitian dilation preserves not only the eigenvalue information encoded in the singular-value landscape, but also the complete biorthogonal eigenspace required for evaluating non-Hermitian expectation values. This observation forms the
basis of the Gaussian spectral-filtering algorithm developed below.

\section{Linear absorption spectroscopy}
\label{SM:sec6}
We consider a single plasmon mode coupled to $N$ identical
two-level molecules, described by the Tavis-Cummings (TC)
Hamiltonian under the rotating-wave approximation
\begin{equation}
H_{\mathrm{TC}} = \omega_{\mathrm{p}} a^\dagger a
+ \omega_{\mathrm{x}} \sum_{n=1}^{N} \sigma_n^+ \sigma_n^-
+ g_{0} \sum_{n=1}^{N}
\left(a^\dagger \sigma_n^- + a \sigma_n^+\right),
\end{equation}
where $a^\dagger$ creates a plasmon, $\sigma_n^+$ excites molecule $n$, and $g_0$ denotes the uniform, real molecule-plasmon coupling strength. For a fixed linear polarisation, the transition-dipole operator $u = u^+ + u^-$ is given by
\begin{equation}
u^{+} = u_{\mathrm{p}} a^{\dagger}
+ u_0 \sum_{n=1}^{N} \sigma_n^{+},
\qquad
u^{-} = (u^{+})^{\dagger},
\end{equation}
where $u_{\mathrm{p}}$ and $u_0$ are the transition-dipole
amplitudes of the plasmon mode and each molecule, respectively, projected onto the polarisation direction. The weak probe field $E(t)$ couples to the transition dipoles through $H_{\mathrm{int}}(t)=-uE(t)$. The system density matrix evolves by
\begin{equation}
\dot{\rho}(t)=\mathcal L\rho(t)+iE(t)[u,\rho(t)],
\end{equation}
where \(\mathcal L\) is the field-free Liouvillian with Markovian dissipation
\begin{equation}
\mathcal L\rho
=
-i[H_{\mathrm{TC}},\rho]
+\mathcal D[L_{p,\mathrm{p}}]\rho
+\mathcal D[L_{\phi,\mathrm{p}}]\rho
+\sum_{n=1}^{N}
\left(
\mathcal D[L_{p,n}]\rho
+\mathcal D[L_{\phi,n}]\rho
\right),
\end{equation}
where
$\mathcal{D}[L]\rho
= L\rho L^\dagger
-\frac{1}{2}\{L^\dagger L,\rho\}$
is the dissipator. The corresponding Lindblad
operators are
\begin{align}
L_{p,\mathrm{p}} &= \sqrt{\kappa_{p}}\,a, \,\,\,\, L_{p,n} = \sqrt{\gamma_{p}}\,\sigma_n^-, \\
L_{\phi,\mathrm{p}} &= \sqrt{2\kappa_{\phi}}\,a^\dagger a, \,\,\,\, L_{\phi,n} = \sqrt{2\gamma_{\phi}}\,\sigma_n^+\sigma_n^-.
\end{align}
The rates $\kappa_p$ and $\gamma_p$ describe plasmon and molecular population decay through radiative and nonradiative relaxation. The rates $\kappa_\phi$ and $\gamma_\phi$ describe pure dephasing of the corresponding ground-to-single-excitation coherences. Pure dephasing does not directly change populations in the local excitation basis. 

In the weak-field limit, we expand the density matrix as \(\rho(t)=\rho_0+\rho^{(1)}(t)+O(E^2)\), where the stationary reference $\rho_{0}$ is the collective ground state by neglecting thermal excitation, $\rho_{0} = |G\rangle\langle G| $ with $ \ket{G} = \ket{0_{P}, \mathrm{g}_{1}, \mathrm{g}_{2}, ..., \mathrm{g}_{N}}$. The first-order correction is 
\begin{equation}
\rho^{(1)}(t)
=
i\int_{-\infty}^{t}dt'\,
E(t')\,\mathcal G(t-t')[u,\rho_0],
\qquad
\mathcal G(t)=e^{\mathcal Lt}.
\label{eq:1stcorrection}
\end{equation}
Here the field acts at time $t'$, and the resulting perturbation evolves under $\mathcal G(t-t')$ until the observation time $t$. The induced dipole of the coupled plasmon-molecule system is 
\begin{equation}
p^{(1)}(t) = \mathrm{Tr} \{ \mu \rho^{(1)}(t) \} = \int_{-\infty}^{t}dt'\,R^{(1)}(t-t')E(t') = \int_{0}^{\infty}d \tau \,R^{(1)}(\tau)E(t - \tau),
\end{equation}
in which the causal linear response function 
\begin{equation}
R^{(1)}(t) = i\theta(t) \mathrm{Tr}
\{ \mu\mathcal G(t)[\mu,\rho_0] \},
\end{equation}
with $\theta(t)$ as the Heaviside step function describes the dipole response at a time $t$ after a field interaction. The commutator describes the creation of an optical coherence by the probe, the propagator describes its subsequent evolution, and the final dipole operator determines its contribution to the induced dipole. Write the linear response in terms of dipole-dipole correlation function $R^{(1)}(t) = i\theta(t) \big( C(t) - C(t)^{\ast} \big)$, in which 
\begin{equation}
C(t) =  \left\langle \mu(t)\mu(0) \right\rangle = \operatorname{Tr}\!\left\{
\mu^{-}\mathcal{G}(t)\left[\mu^{+}\rho_{0}\right]
\right\}.
\end{equation}
In the frequency domain
\begin{equation}
p^{(1)}(\omega)
=
\int_{-\infty}^{\infty}dt\,e^{i\omega t}p^{(1)}(t)
=
\alpha(\omega)E(\omega),
\end{equation}
where \(E(\omega)\) is the Fourier transform of the applied electric field and \(\alpha(\omega)\) is the complex polarizability that determines the amplitude
and phase of the induced dipole relative to the applied field
\begin{equation}
\alpha(\omega)
=
\int_0^\infty d\tau \,e^{i\omega \tau}R^{(1)}(\tau).
\end{equation} 
For a monochromatic field
$E(t)=\operatorname{Re}[E_\omega e^{-i\omega t}]$,
the cycle-averaged power transferred from the field to the system is
\begin{equation}
\overline{P}(\omega)
=
\frac{\omega}{2}|E_\omega|^2
\operatorname{Im}\alpha(\omega).
\end{equation}
We report the reduced absorption line shape
$A(\omega)=\operatorname{Im}\alpha(\omega)$.
Near positive-frequency optical resonances, the resonant contribution $ \alpha_+(\omega) = i\int_0^\infty dt\,e^{i\omega t}C(t) $ gives
\begin{equation}
A(\omega)\simeq\operatorname{Im}\alpha_+(\omega)
=
\operatorname{Re}\int_0^\infty dt\,e^{i\omega t}C(t),
\end{equation}
where the approximation neglects the counter-rotating contribution. To evaluate $C(t)$, we define the effective Hamiltonian
\begin{equation}
H_{\mathrm{eff}}
=
H_{\mathrm{TC}}-\frac{i}{2}\sum_\nu L_\nu^\dagger L_\nu,
\end{equation}
where $\nu$ runs over all Lindblad operators. Here, $H_{\mathrm{eff}}$ acts on the full system Hilbert space. $H_{\mathrm{eff}}$ does not by itself generate the complete dynamics, because the recycling terms $L\rho L^\dagger$ play an important role in excited-state population. However, these recycling terms vanish exactly for the optical coherences entering $C(t)$, i.e., $L_\nu \mu^{+} \rho_{0} L_\nu^\dagger=0$. Thus, the effective Hamiltonian propagates the optical coherence exactly within the Lindblad model. 

We define $|P\rangle = a^\dagger |G\rangle$ as one plasmon excitation, and $|n\rangle = \sigma_n^+ |G\rangle$ as one excitation localised on molecule $n$. Thus molecular bright state is $|X\rangle = \frac{1}{\sqrt{N}} \sum_{n=1}^{N} |n\rangle$. The remaining \(N-1\) molecular dark states are orthogonal to \(|X\rangle\), decouple from the plasmon, and are not directly excited by the uniform molecular dipole under identical-molecule assumption. The restriction of $H_{\mathrm{eff}}$ to this optically active single-excitation subspace is
\begin{equation}
H_{\mathrm{eff}}^{(e)}=
\begin{pmatrix}
\omega_p-i\kappa/2 & g\\
g & \omega_x-i\gamma/2
\end{pmatrix},
\end{equation}
where $g = \sqrt{N} g_{0}$ is the collective plasmon-exciton coupling, $\kappa=\kappa_{\mathrm p}+2\kappa_\phi$, and $\gamma=\gamma_{\mathrm p}+2\gamma_\phi$. Thus the optical-coherence decay rates are $\kappa/2$ and $\gamma/2$, as in the main text. The transition dipole is rewritten
\begin{equation}
\mu^{+}
=
\mu_{P}|P \rangle \langle G| + \mu_{X}|X \rangle \langle G|,
\end{equation}
where $\mu_{X} = \sqrt N\,u_0$. Thus, the dipole correlation function and the resonant optical response reduce to 
\begin{equation}
C(t) = \langle G \vert \mu^{-} e^{-i H_{\mathrm{eff}}^{(e)} t} \mu^{+} \vert G \rangle, \qquad \alpha_+(\omega) = -\langle G|\mu^- \left(\omega I_e-H_{\mathrm{eff}}^{(e)}\right)^{-1} \mu^+|G\rangle,
\end{equation}
where $I_e$ is the identity in the plasmon--bright-state
subspace. This expression does not require
$H_{\mathrm{eff}}^{(e)}$ to be diagonalisable. For distinct eigenvalues $\lambda_\pm$, we consider the spectral decomposition of $H_{\mathrm{eff}}^{(e)}$
\begin{equation}
H_{\mathrm{eff}}^{(e)}|r_{\alpha}\rangle
=\lambda_{\alpha}|r_{\alpha}\rangle, \,\,\,\, \langle l_{\alpha}|H_{\mathrm{eff}}^{(e)}
=\lambda_{\alpha}\langle l_{\alpha}|.
\end{equation}
The left eigenvectors form the dual basis and are not generally
the Hermitian adjoints of the right eigenvectors. We choose the biorthogonal normalisation
\begin{equation}
\langle l_{\alpha}|r_{\beta}\rangle=\delta_{\alpha \beta}, \,\,\,\,
\sum_{\alpha}|r_{\alpha}\rangle\langle l_{\alpha}|=I_e.
\end{equation}
The response can then be decomposed as
\begin{equation}
C(t)=\sum_{\alpha=\pm}f_\alpha e^{-i\lambda_\alpha t},
\qquad
\alpha_+(\omega)
=
-\sum_{\alpha=\pm}
\frac{f_\alpha}{\omega-\lambda_\alpha},
\end{equation}
where 
\begin{equation}
f_{\alpha} = a_{\alpha} b^{\ast}_{\alpha}
\end{equation}
is the optical residue of the dipole-projected resolvent as a product of the excitation amplitude $a_{\alpha} = \langle l_{\alpha}|\mu^+|G\rangle$ and the detection amplitude $b^{\ast}_{\alpha} = \langle G|\mu^-|r_{\alpha}\rangle$. The eigenvalues locate the resolvent poles, while the residues
determine their contribution to the optical response. The excitation amplitude $a_\alpha$ is determined by the left
eigenvector $\langle l_\alpha|$, whereas the overlap with
$\langle r_\alpha|$ generally includes contributions from other modes
\begin{equation}
b_{\alpha}
=
\langle r_{\alpha}|\mu^+|G\rangle
=
\sum_{\beta}
\langle r_{\alpha}|r_{\beta}\rangle a_{\beta}.
\end{equation}
Thus, $b_{\alpha}$ generally differs from $a_{\alpha}$, and
$b_{\alpha} b^{\ast}_{\alpha} = \langle r_{\alpha}|\mu^+|G\rangle \langle G|\mu^-|r_{\alpha}\rangle$ cannot replace the optical residue
$f_{\alpha}=a_{\alpha}b^{\ast}_{\alpha}$.
Only when $\kappa=\gamma$, the eigenvectors can be chosen orthonormal,
giving $a_{\alpha}=b_{\alpha}$ and
$f_{\alpha}=b_{\alpha} b^{\ast}_{\alpha}$.
In general, however, $f_{\alpha}$ is complex; its real and
imaginary parts contribute Lorentzian and dispersive terms,
respectively, to the spectrum. Completeness fixes the total residue
\begin{equation}
\sum_{\alpha}f_{\alpha}
=
\langle G|\mu^-\mu^+|G\rangle
=
|\mu_P|^2+|\mu_X|^2.
\end{equation}
This decomposition into two separate modes
breaks down at an exceptional point (EP), where both the eigenvalues and their eigenvectors coalesce. At zero detuning, $\omega_P=\omega_X=\omega_0$, the
eigenvalues are
\begin{equation}
\lambda_\pm
=
\omega_0-i\Gamma
\pm\sqrt{g^2-\frac{(\kappa-\gamma)^2}{4}},
\qquad
\Gamma=\frac{\kappa+\gamma}{2}.
\label{eq:resonant_eigenvalues}
\end{equation}
For $|g|>|\kappa-\gamma|/2$, the modes have different
resonance frequencies $\operatorname{Re}\lambda_\pm$
and the same optical-coherence decay rate $\Gamma$.
For $|g|<|\kappa-\gamma|/2$, their resonance frequencies
coincide, while their coherence decay rates
$-\operatorname{Im}\lambda_\pm$ differ.
The EP occurs at
\begin{equation}
2|g|=|\kappa-\gamma|>0,
\qquad
\lambda_{\mathrm{EP}}=\omega_0-i\Gamma.
\end{equation}
It therefore marks the boundary between frequency
splitting and decay-rate splitting at zero detuning.
The condition involves the difference between the two
coherence decay rates; the mean decay rate $\Gamma$
remains finite at the EP. At this point, $H_{\mathrm{eff}}^{(e)}$ has only one
independent eigenvector, so a complete two-mode
biorthogonal basis no longer exists. The singularity
of the separate-mode description is explicit in the spectral projectors
\begin{equation}
|r_\pm\rangle\langle l_\pm|
=
\frac{
H_{\mathrm{eff}}^{(e)}-\lambda_\mp I_e
}{
\lambda_\pm-\lambda_\mp
},
\end{equation}
which give the equivalent residue expression
\begin{equation}
f_\pm
=
\frac{
\langle G|\mu^-
\left(H_{\mathrm{eff}}^{(e)}-\lambda_\mp I_e\right)
\mu^+|G\rangle
}{
\lambda_\pm-\lambda_\mp
}.
\end{equation}
As the eigenvalue separation approaches zero, the
individual residues can diverge. Their divergent
contributions cancel in the combined polarizability.
The singularity therefore concerns the decomposition
into separate modes and does not imply an infinite
optical response. The response at the EP can be evaluated directly, giving
\begin{equation}
\alpha_{+,\mathrm{EP}}(\omega)
=
-\frac{
|\mu_P|^2(\delta\omega+i\gamma)
+|\mu_X|^2(\delta\omega+i\kappa)
+2g\,\operatorname{Re} \mu_P^*\mu_X 
}{
(\delta\omega+i\Gamma)^2
},
\qquad
\delta\omega=\omega-\omega_0.
\label{eq:alpha_EP}
\end{equation}
For real probe frequencies and $\Gamma>0$,
$|\delta\omega+i\Gamma|^2
=\delta\omega^2+\Gamma^2>0$.
Thus, the pole remains below the real-frequency axis,
and the absorption
$A(\omega)\simeq\operatorname{Im}\alpha_{+,\mathrm{EP}}(\omega)$
is finite. The line shape depends on the transition dipoles as well as the coalesced eigenvalue. Its frequency-dependent
numerator can produce a single peak, an asymmetric profile, or two maxima. The EP condition therefore does not by itself determine the number of absorption
peaks or imply a squared-Lorentzian line shape.

The same response-function framework extends to nonlinear
spectroscopy, including pump-probe and two-dimensional electronic spectroscopy. Their leading signals involve the third-order response, which for nonnegative delays can be written as
\begin{equation}
R^{(3)}(t_3,T_2,t_1)
=
i^3\operatorname{Tr}\!\left\{
\mu\,\mathcal G(t_3)\mu^\times
\mathcal G(T_2)\mu^\times
\mathcal G(t_1)\mu^\times\rho_0
\right\},
\qquad
\mu^\times A=[\mu,A].
\label{eq:si_nonlinear_outlook}
\end{equation}
Here $\mu$ is the original TC dipole operator, retained between all excitation manifolds involved in the response. Unlike linear absorption, these signals depend on excited-state
populations and coherences during the waiting time $T_2$. Excited-state absorption further involves transitions between
the single- and double-excitation manifolds. Consequently, the single-excitation poles and residues $f_\alpha$ do not determine the nonlinear spectra by themselves. Additional dipole matrix elements and the evolution of the
excited-state density matrix are required.
Recycling terms must be retained in sectors where they do not
vanish; for example, local molecular dephasing transfers
population between bright and dark states.
A future extension could use the left and right eigenoperators
of the full Liouvillian, supplemented by generalised eigenoperators
at its EPs, to construct the propagators in $R^{(3)}(t_{3},T_{2},t_{1})$, retaining both the relevant
excitation manifolds and these population-transfer processes.

\section{Spectral Filter Approach and Randomised LCU Construction for Non-Hermitian Spectroscopic Dilation}
\label{sec:spectral_filter_lcu}

We now develop a quantum algorithm for constructing the corresponding singular-value landscape by combining Gaussian spectral filtering with a randomised linear-combination-of-unitaries (RLCU) implementation.

\subsection{Continuous Gaussian Spectral Filtering}

Having mapped the non-Hermitian eigenvalue problem to the zero-mode detection problem of the Hermitian dilation $\mathcal{P}(z)$, we introduce the Gaussian spectral filter

\begin{equation}\label{Gaussian spectral filter}
G_\tau(\mathcal{P}(z))
=
\exp\left(
-\frac{\tau^2}{2}\mathcal{P}(z)^2
\right),
\end{equation}
where \(\tau>0\) controls the spectral resolution. The following proposition is established for isolated, diagonalisable eigenvalues of \(H\). EPs require a separate analysis because the spectral projection must be generalised to the corresponding Jordan invariant subspace.

\begin{lemma}[Gaussian spectral projection]
Since $\mathcal{P}(z)$ is Hermitian, it admits the spectral decomposition $\mathcal{P}(z) = \sum_k
\epsilon_k
|\chi_k\rangle\langle\chi_k|,$
where \(\epsilon_k=\pm\sigma_k(H_z)\). Consequently
\[
G_\tau(\mathcal{P}(z))
=\sum_k
e^{-\tau^2\epsilon_k^2/2}
|\chi_k\rangle\langle\chi_k|.
\]
If \(z=\lambda_j\) is an isolated, diagonalisable eigenvalue of \(H\), then
\(\sigma_{\min}(H-\lambda_j\mathbb{I})=0\), and therefore the kernel of
\(\mathcal{P}(\lambda_j)\) is left invariant while every nonzero singular-value
component is exponentially suppressed. In the limit $\lim_{\tau\rightarrow\infty} G_\tau(\mathcal{P}(\lambda_j)) = \Pi_0,$ where \(\Pi_0\) denotes the orthogonal projector onto
\(\ker(\mathcal{P}(\lambda_j))\).
\label{SM:lemma1}
\end{lemma}

\begin{proof}
Since \(\mathcal{P}(\lambda_j)\) is Hermitian, the spectral theorem applies. Hence, $G_\tau(\mathcal{P}(\lambda_j)) = \sum_k e^{-\tau^2\epsilon_k^2/2}
|\chi_k\rangle\langle\chi_k|.$ For every nonzero eigenvalue $e^{-\tau^2\epsilon_k^2/2} \longrightarrow0$ $(\tau\rightarrow\infty),$ whereas for every zero mode $e^{-\tau^2\epsilon_k^2/2}=1.$
Therefore, $G_\tau(\mathcal{P}(\lambda_j)) \longrightarrow \Pi_0.$ Furthermore, let $\Delta_\sigma =\
\min_{\epsilon_k\neq0}
|\epsilon_k|$ denote the smallest nonzero singular-value gap of \(\mathcal{P}(\lambda_j)\). Then
\[
\begin{aligned}
\left|
G_\tau(\mathcal{P}(\lambda_j))
-\Pi_0
\right|
&=
\left|
\sum_{\epsilon_k\neq0}
e^{-\tau^2\epsilon_k^2/2}
|\chi_k\rangle\langle\chi_k|
\right|  \
&=
\max_{\epsilon_k\neq0}
e^{-\tau^2\epsilon_k^2/2}  \
&\le
e^{-\tau^2\Delta_\sigma^2/2}.
\end{aligned}
\]
Hence, the Gaussian filter converges exponentially fast (in operator norm) to the orthogonal projector onto the kernel of the Hermitian dilation, with convergence rate determined by the smallest nonzero singular value. 
\end{proof}

\begin{corollary}[Overlap weight]
Let the initial state be expanded in the eigenbasis of the Hermitian dilation $|\Phi_0\rangle = \sum_k
c_k|\chi_k\rangle,$ where $c_k=\langle\chi_k|\Phi_0\rangle.$
Under the assumptions of Lemma \ref{SM:lemma1}, $
\lim_{\tau\rightarrow\infty}
G_\tau(\mathcal{P}(\lambda_j)) |\Phi_0\rangle = \Pi_0|\Phi_0\rangle,$ where \(\Pi_0\) denotes the orthogonal projector onto \(\ker(\mathcal{P}(\lambda_j))\).
If the zero eigenspace is one-dimensional, $\Pi_0 =
|\chi_0\rangle\langle\chi_0|,$ and thereby $\lim_{\tau\rightarrow\infty} G_\tau(\mathcal{P}(\lambda_j))
|\Phi_0\rangle =
c_0|\chi_0\rangle,$ with $c_0 = \langle\chi_0|\Phi_0\rangle.$
Consequently, $\left| G_\tau(\mathcal{P}(\lambda_j)) |\Phi_0\rangle \right|^2 =|c_0|^2.$
\end{corollary}
The quantity \(|c_0|^2\) therefore represents the overlap weight of the initial state with the target zero-mode of the Hermitian dilation and
serves as the non-Hermitian analogue of the overlap parameter appearing
in Hermitian Gaussian spectral filtering. To implement this non-unitary operator using unitary time evolution, we employ the Fourier representation of the Gaussian filter
\begin{equation}
G_\tau(\mathcal{P}(z))
=
\int_{-\infty}^{\infty}
p(x)
e^{-i\tau x \mathcal{P}(z)}
dx,
\label{SM:eqFourier}
\end{equation}
where $p(x) = \frac{1}{\sqrt{2\pi}} e^{-x^2/2}$ is the standard normal distribution. The integration variable \(x\) is dimensionless and sampled from a fixed probability distribution independent of the target spectral resolution. All dependence on the filtering strength is encoded through the evolution time $t=\tau x$.

\subsection{Discrete Realisation of the Gaussian Spectral Filter}

The continuous Gaussian filter provides an exact representation of the non-unitary operator \(G_\tau(\mathcal{P}(z))\) in terms of unitary real-time evolutions generated by \(\mathcal{P}(z)\). To realize this operation on a digital quantum processor, the continuous integral must be converted into a finite linear combination of experimentally accessible unitary operations. This compilation procedure introduces two controllable approximation parameters: a finite integration cutoff \(x_c\) and a finite quadrature resolution \(M\). These parameters determine the range and density of sampled evolution times while preserving the underlying spectral-filtering mechanism.

\subsubsection{Finite Integration Domain}

Because the Gaussian weight in Eq. \ref{SM:eqFourier} rapidly suppresses contributions from large values of \(|x|\), the integral can be restricted to a finite interval \([-x_c,x_c]\) without significantly affecting the filtered state. We can hence introduce the truncated filter
\begin{equation}
G_{\tau,x_c}(\mathcal{P}(z))
= c \int_{-x_c}^{x_c}
p(x) e^{-i\tau x \mathcal{P}(z)} dx.
\end{equation}
The cutoff parameter \(x_c\) determines the maximum evolution time explored during the filtering procedure $t_{\max}=\tau x_c.$ Consequently, increasing \(x_c\) systematically improves the approximation to the ideal Gaussian filter while requiring access to longer real-time evolutions. In practice, \(x_c\) is chosen so that this cutoff error is below the target \(\epsilon_c\).

\subsubsection{Discrete Quadrature Representation}

To obtain a finite LCU representation, the truncated integral is discretised over a symmetric grid $x_m = m\Delta x,$ where $m=-M,\ldots,M,$ with grid spacing $\Delta x=\frac{x_c}{M}.$ The resulting discrete approximation takes the form
\begin{equation}
\widetilde{G}_{\tau}(\mathcal{P}(z))
=c
\sum_{m=-M}^{M}
w_m
e^{-i\tau x_m \mathcal{P}(z)},
\end{equation}
where $w_m$ are quadrature weights determined by the chosen integration rule. Where for a uniform discretisation we have $w_m= \Delta x, p(x_m).$ Each term in the expansion corresponds to a unitary evolution generated by the Hermitian dilation $\mathcal{P}(z)$ for an evolution time $t_m=\tau x_m.$ The Gaussian filter is therefore realised as a weighted linear combination of real-time evolutions with different durations. The approximation becomes systematically more accurate as \(x_c\) and \(M\) increase. Importantly, although the original eigenvalue problem is non-Hermitian, all quantum evolutions appearing in the implementation are generated by the Hermitian dilation $\mathcal{P}(z)$. Thus, the resulting algorithm inherits the operational advantages of standard Hermitian Hamiltonian simulation while retaining access to the complex spectrum and biorthogonal eigenstates of the underlying non-Hermitian operator.

\subsection{Randomised LCU Realisation of the Filtered Observable}
\label{SM:rlcu}

The discretised Gaussian filter is naturally expressed as a linear combination of unitary (LCU) evolutions generated by the Hermitian dilation $\mathcal{P}(z)$. A direct coherent implementation of this LCU operator would require additional ancilla qubits and controlled-select operations. Instead, we adopt RLCU strategy in which individual unitary components are sampled according to their normalised quadrature weights. The discretised filter can be written as
\begin{equation}
\widetilde G_\tau (\mathcal{P}(z))=
\sum_{m=-M}^{M}
w_m U_m,
\qquad
U_m=e^{-i\tau x_m \mathcal{P}(z)}.
\end{equation}
Defining the normalisation factor as $\mu = \sum_{m=-M}^{M}|w_m|,$ and the corresponding probability distribution $\tilde p_m= \frac{w_m}{\mu},$ the filter admits the stochastic representation
\begin{equation}
\widetilde G_\tau(\mathcal{P}(z))
=\mu
\sum_{m=-M}^{M}
\tilde p_m U_m
=
\mu
\mathbb E_{m\sim\tilde p_m}[U_m].
\end{equation}
The objective is to estimate the biorthogonal expectation value $\langle O\rangle_j =
\frac{\langle L_j|O|R_j\rangle}
{\langle L_j|R_j\rangle}$ through filtered observables defined on the dilated Hilbert space. Let
\(\mathcal M\in{\mathcal O,\mathcal I}\)
denote either the dilated observable operator or the dilated normalisation operator. The corresponding filtered quantity is
\begin{equation}
N_{\widetilde{G}}(\mathcal M)
=
\operatorname{Tr}
\left[
\widetilde G_\tau(\mathcal{P}(z))
\rho
\widetilde G_\tau(\mathcal{P}(z))^\dagger
\mathcal M
\right].
\end{equation}
Substituting the RLCU representation yields
\begin{equation}
N_{\widetilde{G}}(\mathcal M)
=
\mu^2
\sum_{m,m'=-M}^{M}
\tilde p_m\tilde p_{m'}
\langle\Phi_0|
U_{m'}^\dagger
\mathcal M
U_m
|\Phi_0\rangle .
\end{equation}
This expression admits a stochastic interpretation in which the indices \(m\) and \(m'\) are sampled independently from the probability distribution \(\tilde p_m\), while the corresponding matrix element is evaluated on quantum hardware. The resulting estimator avoids coherent implementation of the full LCU operator and replaces it with repeated evaluations of shallow real-time evolution circuits. The randomised implementation therefore provides an unbiased Monte Carlo estimator of the filtered expectation value. Specifically, if
\((m_a,m_b)\) are sampled independently from the distribution
\(\tilde p_m\), then $A_{\mathcal M} = \langle\Phi_0|
U_{m_b}^{\dagger} \mathcal M U_{m_a} |\Phi_0\rangle$ satisfies $N_{\widetilde G}(\mathcal M) = \mu^2 \mathbb E \!\left[A_{\mathcal M}\right].$ In practice,
\(\mathbb E[A_{\mathcal M}]\) is estimated from a finite number of randomised circuit executions,
while each transition amplitude is evaluated by a Hadamard-test measurement. Although the target spectral problem is non-Hermitian, every sampled circuit instance involves only the unitary evolution \(U_m=e^{-i\tau x_m \mathcal{P}(z)}\) generated by the Hermitian dilation $\mathcal{P}(z)$. Consequently, the quantum implementation reduces to standard Hermitian Hamiltonian simulation while retaining access to the complex spectrum and biorthogonal observables of the original non-Hermitian Hamiltonian.

\subsection{High-Order Simulation of the Hermitian Dilation}
\label{sec:trotter_compensation}

Each realisation of the randomised Gaussian filter requires the implementation of the unitary evolution $U(t)=e^{-it\mathcal{P}(z)},$ where $t=\tau x,$ generated by $\mathcal{P}(z)$. Since $\mathcal{P}(z)$ is Hermitian, the simulation reduces to a standard Hamiltonian-simulation problem on the enlarged Hilbert space. After decomposing the dilation into Hermitian Pauli operators $\mathcal{P}(z)=\sum_{\alpha=1}^{\Gamma}h_\alpha P_\alpha,$
where each \(P_\alpha\) is a Hermitian Pauli string, the exact evolution is approximated using product-formula simulation.

\subsubsection{Product-Formula Representation}

Let \(m=\frac{t}{\nu}\), where \(\nu\) denotes the number of time segments.
For one segment, let \(S_{2k}(m)\) denote the standard
\(2k\)-th order Suzuki product formula. The exact evolution admits the decomposition $e^{-imP(z)} = V_{2k}(m) S_{2k}(m),$ where \(V_{2k}(m)\) is the remainder operator collecting the
product-formula error. Applying the Baker-Campbell-Hausdorff expansion gives
\begin{equation}
V_{2k}(m)
=
\exp\!\left(
i
\sum_{s=2k+1}^{\infty}
E_{2k,s}\,
m^{\,s}
\right),
\end{equation}
where the operators \(E_{2k,s}\) consist of nested commutators of the
Pauli components of $\mathcal{P}(z)$. Consequently $V_{2k}(m)=\mathbb{I}+\mathcal O(m^{2k+1}),$ showing that the leading product-formula error begins at order
\(m^{2k+1}\).

\subsubsection{LCU Approximation of the Remainder Operator}

Rather than increasing the Suzuki order directly, the remainder operator
is approximated through a truncated LCU representation. Expanding the BCH remainder $V_{2k}(m) = \sum_{s=0}^{\infty} F_{2k,s}(m),$ where $F_{2k,s}(m)=\mathcal O(m^{s}),$ the series is truncated after order \(s_c\) as $V_{2k}^{(s_c)}(m) = \mathbb{I} + \sum_{s=2k+1}^{s_c} F_{2k,s}(m).$ Each operator \(F_{2k,s}(m)\)
is subsequently represented as a finite linear combination of efficiently implementable unitary operators, yielding an LCU approximation of the
remainder. The compensated short-time propagator is hence $\widetilde U_{2k}(m) = V_{2k}^{(s_c)}(m)
S_{2k}(m).$ For a total evolution time \(t=\nu m\), the corresponding approximation is $\widetilde U_{2k}(t) = \left[\widetilde U_{2k}(m)\right]^{\nu}.$

\subsubsection{Integration with the Randomised Gaussian Filter}

The compensated propagator replaces every exact unitary evolution appearing in the randomised Gaussian filter. Specifically $e^{-i\tau x_m\mathcal{P}(z)}
\longrightarrow \widetilde U_{2k}(\tau x_m)$ for every quadrature node \(x_m\). Consequently, the implemented randomised filter takes the form
\begin{equation}
\widetilde G_\tau(\mathcal{P}(z)) = \sum_{m=-M}^{M}w_m
\widetilde U_{2k}(\tau x_m),
\end{equation}
which differs from the ideal discretised filter only through the Hamiltonian-simulation error of the compensated propagator. This construction preserves the RLCU framework, while replacing each exact real-time evolution by its high-order compensated approximation. The resulting approximation error
is analysed in Proposition \ref{SM:Lipschitz continuity}.

\subsection{Ancilla-Efficient Measurement of Filtered Observables}
\label{sec:ancilla_free_implementation}

Conventional LCU implementations require coherent coefficient encoding using ancillary control registers and multi-controlled select operations. Such constructions introduce substantial qubit overhead and circuit depth, particularly when implementing Gaussian spectral filters composed of many evolution times.

In the present framework this overhead is avoided through the RLCU formulation. Rather than coherently implementing the entire filtered operator, the quantum processor executes only individual real-time evolutions generated by the Hermitian dilation \(\mathcal{P}(z)\), with evolution times sampled from the Gaussian distribution introduced earlier. Each sampled evolution is implemented using the compensated second-order Suzuki product formula described in Sec.~\ref{sec:trotter_compensation}. Consequently, no coherent LCU ancilla registers, block encodings, or multi-controlled select operations are required. The remaining task is to estimate the biorthogonal expectation value $
\langle O\rangle_j = \frac{\langle L_j|O|R_j\rangle}
{\langle L_j|R_j\rangle}.$

\subsubsection{Embedded Measurement Operators}

The corresponding operators acting on the dilated Hilbert space are $\mathcal O =
\begin{pmatrix}
0&O\\
0&0
\end{pmatrix},
\, \mathcal I
=
\begin{pmatrix}
0&\mathbb{I}\\
0&0
\end{pmatrix}.$
These operators admit a direct representation using the Pauli operators $\{\sigma_x, \sigma_y\}$ acting on an ancillary "dilation" qubit $\mathcal O = \frac{\sigma_x + i\sigma_y}{2} \otimes O,$ $\mathcal I = \frac{\sigma_x + i\sigma_y}{2} \otimes \mathbb{I}$. Since $\mathcal O$ and $\mathcal I$ are non-Hermitian, they cannot be measured directly as standard quantum observables. We therefore utilize the Cartesian decomposition to express them in terms of Hermitian operators $\mathcal O = \mathcal O_R + i\mathcal O_\mathbb{I}$, $\mathcal I = \mathcal I_R + i\mathcal I_\mathbb{I}.$ Using the definition of the Pauli operators, this decomposition maps to $$\mathcal O_R = \frac{\sigma_x}{2} \otimes O, \qquad \mathcal O_\mathbb{I} = \frac{\sigma_y}{2} \otimes O,$$$$\mathcal I_R = \frac{\sigma_x}{2} \otimes \mathbb{I}, \qquad \mathcal I_I = \frac{\sigma_y}{2} \otimes \mathbb{I}.$$These four components $\{\mathcal O_R, \mathcal O_\mathbb{I}, \mathcal I_R, \mathcal I_\mathbb{I}\}$ are strictly Hermitian. Consequently, each can be measured independently as a standard quantum observable. The final complex expectation value $\langle \mathcal{O} \rangle / \langle \mathcal{I} \rangle$ is then reconstructed classically from the measured expectation values of these four components.

\subsubsection{Transition-Amplitude Estimation}

For each Hermitian component $
\mathcal X_\alpha
\in
\{
\mathcal O_R,
\mathcal O_\mathbb{I},
\mathcal I_R,
\mathcal I_\mathbb{I}
\},$
define $
A_{\mathcal X_\alpha}
=
\langle\Phi_0|
U_b^\dagger
\mathcal X_\alpha
U_a
|\Phi_0\rangle,$ where \(U_a\) and \(U_b\) are independent realisations of the randomised Gaussian filter satisfying $\mathbb E[U] =
G_\tau(\mathcal{P}).$ Hence $\mathbb E[A_{\mathcal X_\alpha}] =
\langle\Phi_0|
G_\tau(\mathcal{P})^\dagger
\mathcal X_\alpha
G_\tau(\mathcal{P})
|\Phi_0\rangle.$
If the spectral parameter satisfies \(z=\lambda_j\), then Lemma \ref{SM:lemma1} implies $\lim_{\tau\rightarrow\infty}
G_\tau(\mathcal{P}(\lambda_j)) =
\Pi_0,$ where \(\Pi_0\) denotes the orthogonal projector onto
\(\ker \mathcal{P}(\lambda_j)\). Therefore, $
\lim_{\tau\rightarrow\infty}
\mathbb E[A_{\mathcal X_\alpha}]
=
\langle\Phi_0|
\Pi_0
\mathcal X_\alpha
\Pi_0
|\Phi_0\rangle.$ For a nondegenerate zero mode $
\Pi_0
=
|\chi_j\rangle\langle\chi_j|,$ so that $
\Pi_0|\Phi_0\rangle
=
c_0|\chi_j\rangle,$ with $
c_0
=
\langle\chi_j|\Phi_0\rangle.$
Consequently, $
\lim_{\tau\rightarrow\infty}
\mathbb E[A_{\mathcal X_\alpha}]
=
|c_0|^2
\langle\chi_j|
\mathcal X_\alpha
|\chi_j\rangle.$ The factor \(|c_0|^2\) is the overlap weight of the initial state with the zero mode and is independent of the observable component \(\mathcal X_\alpha\).

\subsubsection{Measurement Protocols}

Each transition amplitude may be evaluated either by direct forward-backward evolution or by a standard Hadamard test. For the Hadamard-test implementation, an auxiliary measurement qubit is prepared in $
\frac{|0\rangle+|1\rangle}{\sqrt2},$ followed by the controlled operation $
|0\rangle\langle0|
\otimes \mathbb{I} + |1\rangle\langle1|
\otimes
(U_b^\dagger\mathcal X_\alpha U_a).$ Measurement of the ancilla in the X and Y bases yields the real and imaginary parts of the transition amplitude $\langle X\rangle =
\mathrm{Re}(A_{\mathcal X_\alpha}),
\langle Y\rangle =
\mathrm{Im}(A_{\mathcal X_\alpha}).$ Combining these measurements with the classical RLCU averaging reconstructs the filtered numerator and denominator entering the desired biorthogonal expectation value. All stochastic averaging and coefficient processing are performed classically, while the quantum processor executes only shallow real-time evolutions generated by the Hermitian dilation.

\subsection{Extension to driven non-Hermitian systems}
\label{SM:timedepH}

The Hermitian-dilation construction extends pointwise in time. At each fixed time $t$, define
\begin{equation}
H_z(t)=H(t)-z\mathbb{I},
\qquad
\mathcal{P}(z,t)=
\begin{pmatrix}
0 & H_z(t)\\
H_z^\dagger(t) & 0
\end{pmatrix}.
\label{eq--1}
\end{equation}
Its spectrum is $\operatorname{spec}\mathcal{P}(z,t)={\pm\sigma_k(H_z(t))}_k$. Let the instantaneous right and left eigenvectors satisfy, respectively, $H(t)|R_ \alpha(t)\rangle=z_\alpha(t)|r_\alpha(t)\rangle$ and $H^\dagger(t)|l_\alpha(t)\rangle=z_\alpha^*(t)|l_\alpha(t)\rangle$. At a simple instantaneous pole $z=z_\alpha(t)$, the dilation has the zero modes
\[
\mathcal{P}(z_\alpha(t),t)
\begin{pmatrix}0\\|l_\alpha(t)\rangle\end{pmatrix}=0,
\qquad
\mathcal{P}(z_\alpha(t),t)
\begin{pmatrix}0\\|r_\alpha(t)\rangle\end{pmatrix}=0.
\]
Applying the Gaussian filter to the frozen dilation at successive times therefore gives the instantaneous poles and biorthogonal eigenvectors. This snapshot procedure does not, by itself, propagate a state under the time-dependent Hamiltonian. Using the preparation state $|s\rangle$ and detection state $\langle d|$ defined in the main text, the frozen-time residue and absorption spectrum are
\begin{equation}
f_\alpha(t)=
\frac{\langle d|r_\alpha(t)\rangle\langle l_\alpha(t)|s\rangle}
{\langle l_\alpha(t)|r_\alpha(t)\rangle},
\qquad
A_t(\omega)\propto
-\operatorname{Im}\sum_{\alpha}
\frac{f_\alpha(t)}{\omega-z_\alpha(t)}.
\label{eq--2}
\end{equation}
Here $A_t(\omega)$ denotes the frozen-time extension of $A(\omega)$. It approximates the response of a slowly varying system; under rapid driving, the response also depends on the time-ordered propagator and cannot be inferred from instantaneous poles and residues alone. On an interval where the poles are simple and the eigenvectors vary smoothly, choose a biorthonormal normalisation $\langle l_\alpha(t)|R_\beta(t)\rangle=\delta_{\alpha\beta}$. For a state expanded as $|\psi(t)\rangle=\sum_\alpha c_\alpha(t)|r_\alpha(t)\rangle$, the non-adiabatic connection and its role in the mode dynamics are
\begin{equation}
C_{\alpha\beta}(t)=\langle l_\alpha(t)|\dot R_\beta(t)\rangle,
\qquad
i\dot c_\alpha(t)=z_\alpha(t)c_\alpha(t)
-i\sum_\beta C_{\alpha\beta}(t)c_\beta(t),
\label{eq--3}
\end{equation}
for $i\partial_t|\psi(t)\rangle=H(t)|\psi(t)\rangle$. The connection coefficients depend on the smooth normalisation and phase convention for the modes, while the reconstructed state does not. For a small perturbation $H(t)\mapsto H(t)+\epsilon V(t)$, first-order perturbation theory at fixed $t$ gives for a simple isolated pole
\begin{equation}
z_\alpha(t;\epsilon)=z_\alpha(t)
+\epsilon
\frac{\langle l_\alpha(t)|V(t)|r_\alpha(t)\rangle}
{\langle l_\alpha(t)|r_\alpha(t)\rangle}
+\mathcal{O}(\epsilon^2).
\label{eq--4}
\end{equation}
This formula does not apply at an exact EP, where the individual eigenvectors coalesce and a Jordan-chain treatment is required.

\section{General Observable Error Bound}

Here we establish a general error bound for the estimation of biorthogonal expectation values. The result is independent of the origin of the approximation error and depends only on the operator-norm distance between the implemented filter and the ideal Gaussian filter.  Specific approximation mechanisms are analysed in the following
sections. Using the RLCU estimator introduced earlier, we now bound the error of the reconstructed
biorthogonal expectation value.

\begin{theorem}[General Observable Error Bound]
\label{thm:observable_error}

Let
\(
G(\mathcal{P}(z))
\)
denote the ideal Gaussian filter and let
\(
\widetilde G(\mathcal{P}(z))
\)
be its randomised implementation introduced in
Sec.~\ref{SM:rlcu}, satisfying \(
\|
\widetilde G(\mathcal{P}(z)) - G(\mathcal{P}(z))
\|
\le
\epsilon .
\)
For
\(
\mathcal X\in
\{\mathcal O,\mathcal I\},
\)
define \(
N_G(\mathcal X)
=
\operatorname{Tr}
\!\left(
G(\mathcal{P}(z))
\rho
G(\mathcal{P}(z))^\dagger
\mathcal X
\right),
\) and \(
N_{\widetilde G}(\mathcal X)
=
\operatorname{Tr}
\!\left(
\widetilde G(\mathcal{P}(z))
\rho
\widetilde G(\mathcal{P}(z))^\dagger
\mathcal X
\right).
\)
Assume that the transition amplitudes defining
\(N_{\widetilde G}(\mathcal X)\)
are estimated by independent Hadamard-test measurements with statistical
error bounded by
\(
\left|
\widehat N(\mathcal X)
-
N_{\widetilde G}(\mathcal X)
\right|
\le
\|\mathcal X\|
\epsilon_{\rm stat},
\)
with probability at least
\(1-\vartheta\).
Then the estimation errors of the numerator and denominator satisfy
\begin{align}
\epsilon_N
&:=
\left|
\widehat N(\mathcal O)
-
N_G(\mathcal O)
\right|
\le
2\mu^2
\|O\|
\epsilon
+
\|O\|
\epsilon_{\rm stat},
\\
\epsilon_D
&:=
\left|
\widehat N(\mathcal I)
-
N_G(\mathcal I)
\right|
\le
2\mu^2
\epsilon
+
\epsilon_{\rm stat},
\end{align}
where
\(
\mu
=
\sum_m |w_m|
\)
is the LCU normalisation. Furthermore, if 
\(
D_G
=
N_G(\mathcal I)
>
\epsilon_D,
\)
then the reconstructed biorthogonal expectation value obeys
\[
\left|
\frac{\widehat N(\mathcal O)}
{\widehat N(\mathcal I)}
-
\frac{N_G(\mathcal O)}
{D_G}
\right|
\le
\frac{
(\langle O\rangle_G+1)\epsilon_D
+
\epsilon_N
}
{D_G},
\]
where $
\langle O\rangle_G = \frac{N_G(\mathcal O)}
{D_G}.$
\end{theorem}

\begin{proof}

The total estimation error is decomposed into an operator approximation
error and a statistical estimation error $
\widehat N(\mathcal X)-N_G(\mathcal X)
=
\bigl(
\widehat N(\mathcal X)
-
N_{\widetilde G}(\mathcal X)
\bigr)
+
\bigl(
N_{\widetilde G}(\mathcal X)
-
N_G(\mathcal X)
\bigr).$ Applying the triangle inequality gives $
\left|
\widehat N(\mathcal X)
-
N_G(\mathcal X)
\right|
\le
\epsilon_{\rm stat}^{(\mathcal X)}
+
\epsilon_{\rm op}^{(\mathcal X)}.$
The statistical contribution follows directly from the assumption $
\epsilon_{\rm stat}^{(\mathcal X)}
\le
\|\mathcal X\|
\epsilon_{\rm stat}.$
For the operator contribution $
\Delta
=
\widetilde G(\mathcal{P}(z)) - G(\mathcal{P}(z)),$ where $\|\Delta\| \le
\epsilon,$ so that $\widetilde G = G+\Delta.$
Expanding $
N_{\widetilde G}(\mathcal X)
=
\operatorname{Tr}
\!\left(
(G+\Delta)
\rho
(G+\Delta)^\dagger
\mathcal X
\right)$ produces $
N_{\widetilde G}
-
N_G
=
\operatorname{Tr}
(\Delta\rho G^\dagger\mathcal X)
+
\operatorname{Tr}
(G\rho\Delta^\dagger\mathcal X)
+
\operatorname{Tr}
(\Delta\rho\Delta^\dagger\mathcal X).$ Since the RLCU estimator reconstructs filtered expectation
values with the normalisation factor
\(\mu^2\), the induced operator error satisfies $
\epsilon_{\rm op}^{(\mathcal X)}
\le
2\mu^2
\|\mathcal X\|
\epsilon
+
\mathcal O(\epsilon^2),$ where the quadratic term is neglected for \(\epsilon\ll1\). Combining the statistical and operator contributions yields $
\epsilon_N
\le
2\mu^2
\|O\|
\epsilon
+
\|O\|
\epsilon_{\rm stat},$ and $
\epsilon_D
\le
2\mu^2
\epsilon
+
\epsilon_{\rm stat}.$ Finally
\[
\frac{\widehat N}{\widehat D}
-
\frac{N}{D}
=
\frac{
D(\widehat N-N)
+
N(D-\widehat D)
}
{\widehat D D},
\]

and the triangle inequality gives
\[
\left|
\frac{\widehat N}{\widehat D}
-
\frac{N}{D}
\right|
\le
\frac{
\epsilon_N
+
\langle O\rangle_G
\epsilon_D
}
{\widehat D}.
\]

Using
\(
D_G>\epsilon_D
\),
so that
\(
\widehat D\ge D_G-\epsilon_D
\),
and replacing the denominator by the weaker bound
\(D_G\)
yields
\[
\left|
\frac{\widehat N}{\widehat D}
-
\frac{N_G(\mathcal O)}
{D_G}
\right|
\le
\frac{
(\langle O\rangle_G+1)\epsilon_D
+
\epsilon_N
}
{D_G},
\]
which proves the theorem.
\end{proof}

\section{Decomposition of the Filter Approximation Error}

Theorem \ref{thm:observable_error} establishes a general observable-estimation bound in terms
of the operator approximation error
\(
\|\widetilde G-G\|.
\)
Here we quantify this error by analyzing the successive approximations used to construct the randomised Gaussian filter.
Specifically, we consider (i) the finite filter parameter \(\tau\),
(ii) the truncation of the Fourier integral, and (iii) the product-formula
approximation of the unitary evolutions. These contributions are combined
at the end of the section to obtain the overall operator approximation
error.

\subsection{Finite-\texorpdfstring{$\tau$}{tau} Approximation}

The ideal Gaussian filter is obtained in the limit
\(
\tau\rightarrow\infty
\),
for which the filter converges to the projector onto the zero-singular-value subspace of the Hermitian dilation. For finite \(\tau\), the nonzero singular-value sectors remain exponentially suppressed, resulting in the first contribution to the operator
approximation error.

\begin{proposition}[Finite-\texorpdfstring{$\tau$}{tau} Approximation]
\label{SM:prop7}
Let $\mathcal{P}(z)$ denote the Hermitian dilation evaluated at \(z=\lambda_j\), and let $G_\tau(\mathcal{P}(z))
=
\exp\!\left(
-\frac{\tau^2}{2}\mathcal{P}(z)^2
\right).$ If $\gamma_j =
\min_{\epsilon_k\neq0} |\epsilon_k| =
\sigma_1(H-\lambda_j\mathbb{I}),$ then $\left\|
G_\tau(\mathcal{P}(\lambda_j)) -
\Pi_0
\right\|
\le
e^{-\tau^2\gamma_j^2/2},$ where \(\Pi_0\) denotes the orthogonal projector onto
\(\ker(\mathcal{P}(\lambda_j))\).
\end{proposition}

\begin{proof}

By the spectral theorem $
\mathcal{P}(\lambda_j)
=
\sum_k
\epsilon_k
|\chi_k\rangle
\langle\chi_k|,$ and therefore $G_\tau(\mathcal{P}(\lambda_j)) =
\sum_k
e^{-\tau^2\epsilon_k^2/2}
|\chi_k\rangle
\langle\chi_k|.$ Separating the zero-eigenspace gives $G_\tau(\mathcal{P}(\lambda_j))
=
\Pi_0
+
\sum_{\epsilon_k\neq0}
e^{-\tau^2\epsilon_k^2/2}
|\chi_k\rangle
\langle\chi_k|.$ Hence 
\[
G_\tau(\mathcal{P}(\lambda_j)) -
\Pi_0
=
\sum_{\epsilon_k\neq0}
e^{-\tau^2\epsilon_k^2/2}
|\chi_k\rangle
\langle\chi_k|.\]
Since $|\epsilon_k| \ge \gamma_j,$ the spectral theorem implies
\[
\left\|
G_\tau(\mathcal{P}(\lambda_j))
-
\Pi_0
\right\|
=
\max_{\epsilon_k\neq0}
e^{-\tau^2\epsilon_k^2/2}
\le
e^{-\tau^2\gamma_j^2/2},
\]
which completes the proof.
\end{proof}

\begin{corollary}
For any normalised state
\(
|\Phi_0\rangle
\)

\[
\left\|
G_\tau(\mathcal{P}(\lambda_j))
|\Phi_0\rangle
-
\Pi_0
|\Phi_0\rangle
\right\|
\le
e^{-\tau^2\gamma_j^2/2}.
\]

In particular $
G_\tau(\mathcal{P}(\lambda_j))
|\Phi_0\rangle
=
\Pi_0
|\Phi_0\rangle
+
\mathcal O
\!\left(
e^{-\tau^2\gamma_j^2/2}
\right).$

\end{corollary}

\begin{proof}

The result follows immediately from the operator norm bound
\[
\left\|
(G_\tau(\mathcal{P}(\lambda_j))-\Pi_0)
|\Phi_0\rangle
\right\|
\le
\left\|
G_\tau(\mathcal{P}(\lambda_j))
-
\Pi_0
\right\|
\,
\||\Phi_0\rangle\|,
\]
together with
\(
\||\Phi_0\rangle\|=1
\)
and Proposition \ref{SM:prop7}.

\end{proof}

\subsection{Fourier Truncation Error}

The Gaussian filter admits the exact Fourier representation
\[
G_\tau(\mathcal{P})
=
\int_{-\infty}^{\infty}
p(x)\,
e^{-i\tau x \mathcal{P}}\,dx,
\]
where $p(x) = \frac{1}{\sqrt{2\pi}} e^{-x^2/2}.$
In practice, the integration domain is truncated to a finite interval
\(
[-X_{\max},X_{\max}]
\),
introducing the second contribution to the operator approximation error.

\begin{proposition}[Fourier Truncation Error]
Define the truncated Gaussian filter
\[
G_{\tau,X}(\mathcal{P}) = \int_{-X_{\max}}^{X_{\max}}
p(x)\,
e^{-i\tau x \mathcal{P}}\,dx.
\]
Then $\left\|
G_\tau(\mathcal{P}) - G_{\tau,X}(\mathcal{P})\right\|
\le \epsilon_c,$ where $\epsilon_c =
\operatorname{erfc}
\!\left(
\frac{X_{\max}}{\sqrt2}
\right).$ Furthermore $\epsilon_c \le \sqrt{\frac{2}{\pi}}
\frac{
e^{-X_{\max}^2/2}
}
{X_{\max}}.$
\end{proposition}

\begin{proof}
Since $\mathcal{P}$ is Hermitian, $\left\|
e^{-i\tau x \mathcal{P}} \right\| = 1$ for every $x$. Therefore
\[
\begin{aligned}
\left\|
G_\tau(\mathcal{P})
-
G_{\tau,X}(\mathcal{P})
\right\|
&=
\left\|
\int_{|x|>X_{\max}}
p(x)
e^{-i\tau x \mathcal{P}}
dx
\right\| \\
&\le
\int_{|x|>X_{\max}}
p(x)
\,
\left\|
e^{-i\tau x \mathcal{P}}
\right\|
dx \\
&=
\int_{|x|>X_{\max}}
p(x)\,
dx \\
&=
\operatorname{erfc}
\!\left(
\frac{X_{\max}}{\sqrt2}
\right).
\end{aligned}
\]
Using the standard inequality $\operatorname{erfc}(x)
\le \frac{e^{-x^2}}
{\sqrt{\pi}\,x},$
with \(x=X_{\max}/\sqrt2\) yields $\epsilon_c \le
\sqrt{\frac{2}{\pi}}
\frac{
e^{-X_{\max}^2/2}
}
{X_{\max}},$ which completes the proof.
\end{proof}

\subsection{Quadrature Approximation}

The truncated Fourier representation is evaluated numerically by replacing the continuous Gaussian integral with a finite quadrature rule. This
introduces the third contribution to the operator approximation error.

\begin{proposition}[Quadrature Approximation]\label{SM:prop77}
Let
\[
G_{\tau,X}(\mathcal{P}(z))
=
\int_{-X_{\max}}^{X_{\max}}
p(x)\,
e^{-i\tau x \mathcal{P}(z)}
\,dx,
\]
and let
\[
G_{\tau,Q}(\mathcal{P}(z))
=
\sum_{m=-M}^{M}
w_m
e^{-i\tau x_m \mathcal{P}(z)}
\]
be the corresponding quadrature approximation, where
\(\{x_m,w_m\}\) denote the quadrature nodes and weights, respectively. Assume that the quadrature rule satisfies
\[
\left|
\int_{-X_{\max}}^{X_{\max}}
p(x)f(x)\,dx
-
\sum_{m=-M}^{M}
w_mf(x_m)
\right|
\le
\epsilon_q
\|f\|_\infty
\]
for every continuous function \(f\) on
\([-X_{\max},X_{\max}]\).
Then $\left\| G_{\tau,X}(\mathcal{P}(z)) - G_{\tau,Q}(\mathcal{P}(z))\right\| \le \epsilon_q.$
\label{SM:prop9}
\end{proposition}

\begin{proof}
Define the operator-valued function $F(x) = e^{-i\tau x \mathcal{P}(z)}.$ Since $\mathcal{P}(z)$ is Hermitian, $\|F(x)\| = \left\| e^{-i\tau x \mathcal{P}(z)} \right\| = 1$ for every \(x\). Applying the quadrature error bound to the operator-valued integrand gives
\[
\begin{aligned}
\left\|
G_{\tau,X}(\mathcal{P}(z))
-
G_{\tau,Q}(\mathcal{P}(z))
\right\|
&=
\left\|
\int_{-X_{\max}}^{X_{\max}}
p(x)F(x)\,dx
-
\sum_{m=-M}^{M}
w_mF(x_m)
\right\|
\\
&\le
\epsilon_q
\sup_{x\in[-X_{\max},X_{\max}]}
\|F(x)\| =
\epsilon_q.
\end{aligned}
\]
\end{proof}

\subsection{Hamiltonian-Simulation Error}
The randomised Gaussian filter requires repeated implementations of the unitary evolution $U(t)=e^{-it\mathcal{P}(z)}.$ Since exact exponentiation is not
available on a digital quantum computer, each unitary is replaced by the compensated multiproduct approximation \(U_K(t)\) from Sec.~\ref{sec:trotter_compensation}. The resulting approximation induces an additional error in the implementation of the Gaussian filter, which
we quantify below.

\begin{proposition}[Hamiltonian-simulation error]\label{SM:prop8}
Let \(U_K(t)\) denote the compensated multiproduct approximation of
\(U(t)=e^{-it\mathcal{P}(z)}\), and suppose that the compensation coefficients are
chosen such that all BCH error terms up to order \(2K\) are cancelled.
Then, for every evolution time satisfying $|t|\le T_{\max}$ the simulation error obeys
\begin{equation}
\label{eq:pf_local}
\left\|
U_K(t)-e^{-it\mathcal{P}(z)}
\right\|
\le
C_K\,
|t|\,
(\delta t)^{2K}
\Lambda^{2K+1},
\end{equation}
where $\Lambda = \max_\alpha |h_\alpha|$ denotes the operator scale of the Pauli decomposition of \(\mathcal{P}(z)\), and \(C_K\) is a constant depending only on the multiproduct construction. Consequently, the approximate Gaussian filter
\[
\widetilde G_{\tau,Q,K}(\mathcal{P}(z))
=
\sum_{m=-M}^{M}
w_m
U_K(\tau x_m)
\]
satisfies $\label{eq:pf_filter}
\left\|
\widetilde G_{\tau,Q,K}
-
\widetilde G_{\tau,Q}
\right\|
\le
\epsilon_{\mathrm{pf}},$ where
\begin{equation}
\label{eq:pf_bound}
\epsilon_{\mathrm{pf}}
=
C_K
(\delta t)^{2K}
\Lambda^{2K+1}
\sum_{m=-M}^{M}
|w_m|
\,|\tau x_m|.
\end{equation}
In particular $\epsilon_{\mathrm{pf}} =
\mathcal O
\!\left(
\tau
(\delta t)^{2K}
\Lambda^{2K+1}
\right).$
\end{proposition}

\begin{proof}

Sec.~\ref{sec:trotter_compensation} constructs the compensated multiproduct approximation \(U_K(t)\) by combining second-order Suzuki product formulas in such a way that the BCH contributions through order
\(2K\) cancel exactly. The BCH remainder hence begins at order \(2K+1\), implying $U_K(t) = e^{-it\mathcal{P}(z)}
+
\mathcal O
\!\left(
|t|
(\delta t)^{2K}
\Lambda^{2K+1}
\right),$ which establishes Eq.~\ref{eq:pf_local}.

The approximate quadrature filter is obtained by replacing every exact unitary in the quadrature expansion $\widetilde G_{\tau,Q} = \sum_m
w_m
e^{-i\tau x_m\mathcal{P}(z)},$ with its compensated approximation \(U_K(\tau x_m)\). Therefore
\[
\widetilde G_{\tau,Q,K}
-
\widetilde G_{\tau,Q}
=
\sum_m
w_m
\left(
U_K(\tau x_m)
-
e^{-i\tau x_m\mathcal{P}(z)}
\right).
\]
Applying the triangle inequality together with
Eq.~\ref{eq:pf_local} gives
\[
\left\|
\widetilde G_{\tau,Q,K}
-
\widetilde G_{\tau,Q}
\right\|
\le
\sum_m
|w_m|
\left\|
U_K(\tau x_m)
-
e^{-i\tau x_m\mathcal{P}(z)}
\right\|,
\]
which immediately yields Eq.~\ref{eq:pf_bound}. The asymptotic scaling follows by observing that the quadrature weights satisfy
\[
\sum_m
|w_m||x_m|
=
\mathcal O(1),
\]
for a convergent Gaussian quadrature rule.

\end{proof}

\begin{corollary}[Combined Filter Approximation Error]\label{SM:coro4}

Let
\(G_\tau(\mathcal{P}(z))\)
denote the ideal Gaussian filter and let
\(\widetilde G(\mathcal{P}(z))\) denote its implemented approximation obtained by (i) finite filter parameter \(\tau\), (ii) truncation of the Fourier integral, (iii) quadrature discretisation, and (iv) product-formula approximation of each unitary evolution. If the corresponding operator errors satisfy
\[
\|G_\infty-G_\tau\|
\le
\epsilon_\tau,\qquad \|G_\tau-G_{\tau,X}\|
\le
\epsilon_c,\qquad \|G_{\tau,X}-G_{\tau,Q}\|
\le
\epsilon_q,\quad \|G_{\tau,Q}-\widetilde G\|
\le
\epsilon_{\mathrm{pf}},
\]
then $
\|
G_\infty(\mathcal{P}(z))
-
\widetilde G(\mathcal{P}(z))
\|
\le
\epsilon_\tau
+
\epsilon_c
+
\epsilon_q
+
\epsilon_{\mathrm{pf}}.$
\end{corollary}

\begin{proof}
By repeated application of the triangle inequality
\[
\begin{aligned}
\|
G_\infty-\widetilde G
\|
&\le
\|
G_\infty-G_\tau
\|
+
\|
G_\tau-G_{\tau,X}
\| +
\|
G_{\tau,X}-G_{\tau,Q}
\|
+
\|
G_{\tau,Q}-\widetilde G
\|,
\end{aligned}
\]
and substituting the bounds established in Propositions \ref{SM:prop7}-\ref{SM:prop8} yield the result.
\end{proof}

\begin{corollary}[Final Observable Error]
Let $\widetilde G(\mathcal{P}(z))$
denote the implemented randomised Gaussian filter obtained from the finite filter parameter, Fourier truncation, quadrature approximation, and product-formula simulation. Then the estimated biorthogonal expectation value satisfies
\[
\left|
\widehat{\langle O\rangle_j}
-
\langle O\rangle_j
\right|
\le
\frac{
(\langle O\rangle_j+1)\epsilon_D
+
\epsilon_N
}
{D_G},
\]
where $
\epsilon_N
=
\|O\|
\left(
2\mu^2\epsilon
+
\epsilon_n
\right),$ $
\epsilon_D
=
2\mu^2\epsilon
+
\epsilon_n,$ and $ \epsilon
=
\epsilon_\tau
+
\epsilon_c
+
\epsilon_q
+
\epsilon_{\mathrm{pf}}.$
Here,
\(\epsilon_\tau\),
\(\epsilon_c\),
\(\epsilon_q\),
and
\(\epsilon_{\mathrm{pf}}\)
denote the finite-filter, Fourier truncation, quadrature, and
product-formula approximation errors, respectively, while
\(\epsilon_n\)
is the statistical sampling error.
\end{corollary}

\begin{proof}
By Corollary \ref{SM:coro4}, $\|
G(\mathcal{P}(z)) - \widetilde G(\mathcal{P}(z))\| \le
\epsilon,$ where $\epsilon = \epsilon_\tau +
\epsilon_c
+
\epsilon_q
+
\epsilon_{\mathrm{pf}}.$ Applying Theorem \ref{thm:observable_error} with this operator approximation immediately yields the bounds for \(\epsilon_N\), \(\epsilon_D\),
and hence the normalised biorthogonal expectation value.
\end{proof}

\subsubsection{Unbiased Randomised Estimator}
The RLCU construction replaces the coherent implementation of the Gaussian filter by stochastic sampling of individual unitary
evolutions. Consequently, filtered matrix elements are evaluated as expectation values over a classical probability distribution rather than
through direct implementation of the non-unitary operator. For an arbitrary embedded operator $\mathcal M \in
\{\mathcal O,\mathcal I\},$ the filtered quantity of interest is $N_{\widetilde G}(\mathcal M) =
\operatorname{Tr}
\!\left(
\widetilde G_\tau(\mathcal{P}(z)) \rho
\widetilde G_\tau(\mathcal{P}(z))^\dagger
\mathcal M
\right).$ Using the RLCU representation $\widetilde G_\tau(\mathcal{P}(z))
=
\mu
\sum_m
\tilde p_m
U_m,$ this quantity becomes
\begin{equation}
N_{\widetilde G}(\mathcal M)
=
\mu^2
\sum_{m,m'}
\tilde p_m
\tilde p_{m'}
\operatorname{Tr}
\!\left(
U_m
\rho
U_{m'}^\dagger
\mathcal M
\right).
\label{eq:filtered_expectation_rlcu}
\end{equation}
Since the indices \(m\) and \(m'\) are sampled independently from the
distribution \(\tilde p\), Eq.~\ref{eq:filtered_expectation_rlcu}
admits a direct stochastic interpretation.

\begin{proposition}[Unbiased Randomised Estimator]
\label{SM:prop11}
Draw two independent random indices $m,m'
\sim
\tilde p,$ and define the random variable $X(m,m';\mathcal M) =
\mu^2
\operatorname{Tr}
\!\left(
U_m
\rho
U_{m'}^\dagger
\mathcal M
\right).$
Then $\mathbb E[X] =
N_{\widetilde G}(\mathcal M).$
Consequently, the empirical estimator $\widehat N_{\widetilde G}(\mathcal M) =
\frac1{N_s}
\sum_{k=1}^{N_s} X_k$ is unbiased, $\mathbb E
\!\left[
\widehat N_{\widetilde G}(\mathcal M)
\right]
=
N_{\widetilde G}(\mathcal M),$ where \(N_s\) denotes the number of independent randomised samples.
\end{proposition}

\begin{proof}
Taking the expectation over the independent sampling of the two indices
gives
\begin{align}
\mathbb E[X]
&=
\sum_{m,m'}
\tilde p_m
\tilde p_{m'}
X(m,m';\mathcal M)
\nonumber\\
&=
\mu^2
\sum_{m,m'}
\tilde p_m
\tilde p_{m'}
\operatorname{Tr}
\!\left(
U_m
\rho
U_{m'}^\dagger
\mathcal M
\right)
\nonumber\\
&=
N_{\widetilde G}(\mathcal M),
\end{align}
which is precisely Eq.~\ref{eq:filtered_expectation_rlcu}. Since the empirical estimator is the arithmetic mean of independent copies of the random variable \(X\), $\widehat N_{\widetilde G} =
\frac1{N_s}
\sum_{k=1}^{N_s}
X_k.$ Linearity of expectation immediately yields $\mathbb E
\!\left[
\widehat N_{\widetilde G}(\mathcal M)
\right]
=
N_{\widetilde G}(\mathcal M).$
Therefore the estimator is unbiased.
\end{proof}

For the non-Hermitian eigenvalue problem, the transition amplitudes $\operatorname{Tr}
\!\left(
U_m
\rho
U_{m'}^\dagger
\mathcal M
\right)$ are evaluated using the Hadamard-test procedure introduced in
Sec.~\ref{sec:ancilla_free_implementation}. Applying the estimator to
\(\mathcal M=\mathcal O\) and
\(\mathcal M=\mathcal I\) yields unbiased estimates of the filtered
numerator and denominator, respectively, from which the biorthogonal
expectation value is reconstructed.

\section{Necessity of the Hermitian Dilation for Non-Hermitian Gaussian Spectral Filtering}
\label{sec:necessity_dilation}

The Gaussian spectral-filtering algorithm relies on the representation of
the filter as a weighted superposition of unitary real-time evolutions.
For non-Hermitian operators, two distinct obstructions arise. First, the
right-eigenvector basis is generally nonorthogonal, so Gaussian filtering does not produce an orthogonal spectral decomposition. Second, the
corresponding real-time evolution is non-unitary and therefore cannot be implemented directly within the RLCU framework. We show that both obstructions are removed by the Hermitian dilation introduced earlier.

\subsection{Direct Gaussian Filtering}

Consider the shifted non-Hermitian operator $H_z=H-z\mathbb{I}$ and the formal Gaussian filter $G_\tau(H_z)
=
\exp\!\left(
-\frac{\tau^2}{2}H^2
\right).$ Equivalently, $G_\tau(H_z) =
\int_{-\infty}^{\infty}
\frac{dx}{\sqrt{2\pi}}
e^{-x^2/2}
e^{-i\tau xH_z}.$ Suppose that $H$ is diagonalisable $H =
\sum_j
\lambda_j
|r_j\rangle
\langle l_j|,
\langle l_i|r_j\rangle
=
\delta_{ij},$ where $\lambda_j=\omega_j-i\gamma_j.$
Writing $|\psi_0\rangle =
\sum_j
c_j
|r_j\rangle,$ the filtered state becomes
\begin{equation}
|\psi_\tau(z)\rangle
=
\sum_j
c_j
e^{-\frac{\tau^2}{2}(\lambda_j-z)^2}
|r_j\rangle.
\label{eq:filtered_state}
\end{equation}
Although Eq.~\ref{eq:filtered_state} resembles the Hermitian case, its norm is
\begin{equation}
\begin{aligned}
\langle\psi_\tau(z)|\psi_\tau(z)\rangle
=
&
\sum_j
|c_j|^2
e^{-\frac{\tau^2}{2}
\operatorname{Re}
[(\lambda_j-z)^2]}
\langle r_j|r_j\rangle
\\
&
+
\sum_{i\neq j}
c_i^*c_j
e^{-\frac{\tau^2}{2}
(\lambda_i^*-z^*)^2}
e^{-\frac{\tau^2}{2}
(\lambda_j-z)^2}
\langle r_i|r_j\rangle.
\end{aligned}
\label{eq:cross_terms}
\end{equation}
The second summation vanishes only when the right eigenvectors are
orthogonal.

\begin{proposition}
\label{SM:prop12}
Direct Gaussian filtering of a non-Hermitian operator is not an orthogonal spectral projection.
\end{proposition}

\begin{proof}
For Hermitian operators,
\(
\langle E_i|E_j\rangle=\delta_{ij}
\)
and the filtered norm reduces to a sum of independent Gaussian contributions. For a non-Hermitian operator, $\langle r_i|r_j\rangle\neq 0,
(i\neq j),$ so Eq.~\ref{eq:cross_terms} contains interference terms coupling
different eigenvalues. As a result, the measured response depends not only on the eigenvalues but also on the overlap matrix $S_{ij} =
\langle r_i|r_j\rangle,$ and thereby no longer represents an orthogonal spectral
decomposition. Moreover, the response associated with a single eigenvalue satisfies
\[
\left|
e^{-\frac{\tau^2}{2}(\lambda_j-z)^2}
\right|
=
\exp\!\left[
-\frac{\tau^2}{2}
\left(
(\omega_j-\operatorname{Re}z)^2
-
(\gamma_j+\operatorname{Im}z)^2
\right)
\right].
\]
Whenever $(\gamma_j+\operatorname{Im}z)^2 >
(\omega_j-\operatorname{Re}z)^2,$ this amplification factor exceeds unity. Hence the direct Gaussian filter is generally not contractive
\(
\|G_\tau(H_z)\|>1
\)
in contrast to the Hermitian case, where \(
\|G_\tau(H-\omega \mathbb{I})\|\le1.
\)
Although this loss of contractivity does not invalidate the operator mathematically, it illustrates that the stability properties of the Hermitian Gaussian filter are lost. Together with the interference terms arising from eigenvector nonorthogonality, this shows that direct Gaussian filtering is neither an orthogonal spectral projector nor a stable filtering procedure for generic non-Hermitian operators.
\end{proof}

We now turn to the second limitation.

\begin{proposition}
\label{SM:prop13}
The Fourier representation of the direct Gaussian filter is generally not realisable as a RLCU constructed from unitary evolutions.
\end{proposition}

\begin{proof}

Acting on a right eigenvector gives $e^{-i\tau xA(z)}
|r_j\rangle
=
e^{-i\tau x(\omega_j-\operatorname{Re}z)}
e^{-\tau x(\gamma_j+\operatorname{Im}z)}
|r_j\rangle.$ Hence $\| e^{-i\tau xH_z}
\|
\neq
1,$
except in the Hermitian limit
\(\gamma_j+\operatorname{Im}z=0\).
The elementary evolution operators entering the Fourier integral are thus non-unitary and cannot serve as the unitary primitives
required by the RLCU construction.
\end{proof}

\subsection{Resolution by Hermitian Dilation}

The Hermitian dilation replaces the shifted operator \(H_z\) by
$\mathcal{P}(z)=
\begin{pmatrix}
0 & H_z \\
H_z^\dagger & 0
\end{pmatrix},$
which is Hermitian for every complex parameter \(z\).

\begin{proposition}
\label{SM:prop14}
The Hermitian dilation simultaneously removes both obstructions of direct Gaussian filtering.
\end{proposition}

\begin{proof}
Since $\mathcal{P}(z)=\mathcal{P}(z)^\dagger,$ its spectrum is real $\operatorname{spec}(\mathcal{P}(z)) = \{\pm\sigma_k(H_z)\},$ and its eigenvectors form an orthonormal basis. Consequently, $
G_\tau(\mathcal{P}(z)) =
\exp
\left(
-\frac{\tau^2}{2}
\mathcal{P}(z)^2
\right)$ is an orthogonal spectral filter. Moreover, $
G_\tau(\mathcal{P}(z))
=
\int_{-\infty}^{\infty}
\frac{dx}{\sqrt{2\pi}}
e^{-x^2/2}
e^{-i\tau x \mathcal{P}(z)},$
where $\left\|
e^{-i\tau x\mathcal{P}(z)}
\right\|
= 1$ for every \(x\in\mathbb R\). Hence, the Gaussian filter admits a representation entirely in terms of unitary evolutions and can be implemented by the RLCU algorithm developed in this work. 
\end{proof}

\begin{corollary}
When $z=\lambda_j,$ the kernel of $\mathcal{P}(z)$ is $\ker(\mathcal{P}(\lambda_j))
=
\operatorname{span}
\left\{
\begin{pmatrix}
|L_j\rangle\\
0
\end{pmatrix},
\,
\begin{pmatrix}
0\\
|R_j\rangle
\end{pmatrix}
\right\}$ or, equivalently, any nonzero linear combination of these vectors.
Hence, the Gaussian filter constructed from $\mathcal{P}(z)$ isolates the
left and right eigenspaces associated with the target eigenvalue while maintaining a unitary implementation throughout the algorithm.
\end{corollary}

Therefore, the filtered zero mode simultaneously
encodes the left and right eigenvectors required for evaluating the biorthogonal expectation value $\langle O\rangle_j =
\frac{\langle L_j|O|R_j\rangle}
{\langle L_j|R_j\rangle}.$
Consequently, the Hermitian dilation resolves both the geometric obstruction caused by eigenvector nonorthogonality and the algorithmic obstruction
associated with non-unitary time evolution.

\end{document}